\documentclass[twocolumn,prd,amsmath,amssymb,epsf,superscriptaddress]{revtex4-1}

\usepackage[dvipdfmx]{graphicx}
\usepackage{dcolumn}
\usepackage{bm}
\usepackage{mathrsfs}
\usepackage{float}
\usepackage{color}
\usepackage{orcidlink}

\begin{document}


\title{Cosmic-ray  acceleration in core-collapse supernova remnants with the wind termination shock}
\author{Shoma F. Kamijima}
\affiliation{Yukawa Institute for Theoretical Physics, Kyoto University, Oiwake-cho, Sakyo-ku, Kyoto-city, Kyoto, 606-8502, Japan}
\affiliation{Department of Earth and Planetary Science, The University of Tokyo, 7-3-1 Hongo, Bunkyo-ku, Tokyo 113-0033, Japan}
\author{Yutaka Ohira}
\affiliation{Department of Earth and Planetary Science, The University of Tokyo, 7-3-1 Hongo, Bunkyo-ku, Tokyo 113-0033, Japan}


\begin{abstract}
We investigate the attainable maximum energy of particles accelerated in the core-collapse supernova remnant (SNR) shock propagating in the free wind region with the Parker-spiral magnetic field, current sheet, and the wind termination shock (WTS) by using test particle simulations.
This work focuses on Wolf-Rayet stars as progenitors.
The magnetic field amplification in the free wind region (shock upstream region) is not considered in this work.
Test particle simulations show that particles escaped from the core-collapse SNR reach and move along the WTS, and eventually return to the SNR shock from the poles or equator of the WTS.
The particle attainable energy can be boosted by this cyclic motion between the SNR shock and WTS and can be larger than the particle energy that is limited by escape from the SNR shock.
The particle energy limited by the cyclic motion between the SNR shock and WTS is about $10$-$100~{\rm TeV}$.
Thus, the core-collapse SNR without upstream magnetic field amplification can be the origin of the break around $10~{\rm TeV}$ of the energy spectrum of observed cosmic ray protons and helium.
\end{abstract}

\maketitle

\section{Introduction} \label{sec:intro}
The origin of cosmic rays (CRs) has been a longstanding problem since the discovery of CRs in 1912.
It is believed that CRs below $3~{\rm PeV}$ are accelerated by the diffusive shock acceleration (DSA) in Galactic supernova remnants (SNRs).
The gamma rays above $100~{\rm TeV}$ from Galactic objects were recently observed, which would lead to revealing the origin of PeV CRs \citep{3pev}.
Future observations in the southern hemisphere (ALPACA \citep{alpaca} and SWGO \citep{swgo}) advance to the elucidation of the origin of PeV CRs.
On the other hand, recent observations reported that the energy spectrum of CR protons and helium has a new spectral break around $10~{\rm TeV}$.
Three ${\rm PeV}$ is thought to be the maximum energy scale of protons originating from Galactic objects. 
On the other hand, it is still unclear what the energy scale of $10~{\rm TeV}$ means; nevertheless, some models for explaining the $10~{\rm TeV}$ break are proposed \citep{10tev,10tevth,kamijima21,kamijima22}.

In the DSA, particles perform the back-and-forth motion across the shock front many times.
These diffusive particles gain energy by the relative motion between the upstream and downstream regions \citep{cr}.
The acceleration time of the DSA depends on the angle between the magnetic field and the shock normal direction \citep{drury83}.
In parallel shocks where the magnetic field is parallel to the shock normal direction, the magnetic field in the shock upstream region has to be amplified to 100 times the typical interstellar magnetic field strength to accelerate particles to the PeV scale \citep{cesarsky81}.
It is still unclear which magnetic field amplification mechanism works in SNRs although some magnetic field amplification mechanisms are proposed \citep{crsi}.
Contrary to parallel shocks, in perpendicular shocks where the magnetic field is perpendicular to the shock normal direction, it is proposed that particles can be accelerated up to the PeV scale without the magnetic field amplification in the shock upstream region \citep{jokipii87}.
This is originated from the rapid acceleration induced by the gyration \citep{takamoto15,kamijima20}.
This rapid acceleration induced by the gyration is shown by numerical simulations \citep{rapidperp,kamijima20}.
The shock geometry plays an important role for acceleration \citep{bell08} although the above work considered the plane shock.

It is shown that not only acceleration but also escape from systems determines the maximum attainable energy of particles and the energy spectrum of observed CRs \citep{ptuskin03, ohira10}.
To investigate the perpendicular shock acceleration and escape process from perpendicular shocks, we have to treat the global particle motion in the whole system while solving the gyration.
In our recent work, we considered the spherical SNR shock in the interstellar medium (ISM) and circumstellar medium (CSM) magnetic fields and performed global test particle simulations that solve the gyration in the ISM and CSM magnetic fields to reveal the escape process from perpendicular shocks \citep{kamijima21,kamijima22}.
For the case of type Ia SNRs in the uniform ISM magnetic field, we showed that the maximum attainable energy, which is about $10~{\rm TeV}$ for CR protons, is limited by escape from the perpendicular shock region \citep{kamijima21}.
As for core-collapse SNRs, we took into account the Parker-spiral magnetic field and the current sheet created by the stellar wind of massive stars (progenitors).
For the case of core-collapse SNRs without upstream magnetic field amplifications, we showed that the maximum attainable energy is limited by escape from the polar or equatorial regions of core-collapse SNRs and the typical escape-limited maximum energy is about $10~{\rm TeV}$ for CR protons \citep{kamijima22}.

Particles escaped from core-collapse SNRs reach the wind termination shock (WTS) created by the stellar wind of progenitors before the supernova explosion.
WTSs are thought to be created by the wind in various systems (e.g. solar wind \citep{solar}, massive star wind \citep{massive,weaver77}, star cluster \citep{cluster}, pulsar wind \citep{pulsar}, and galactic wind \citep{galactic}) and are expected to make important roles for the particle acceleration.
The WTS of the Wolf-Rayet (WR) star is suggested to accelerate electrons by the observation \citep{prajapati19}.
However, our previous study did not take account of the WTS. 
In this work, we investigate the particle dynamics and maximum attainable energy of accelerated particles by using test particle simulations in the entire global system that includes both the core-collapse SNR shock and WTS.
The system and simulation setups we consider are shown in Sec.~\ref{sec:setup}.
In Sec.~\ref{sec:theory}, we discuss the expected particle dynamics between the SNR shock and WTS and the theoretical estimate of the attainable maximum energy of accelerated particles.
The simulation results are shown in Sec.~\ref{sec:result}.
Sections~\ref{sec:discuss} and \ref{sec:summary} are devoted to a discussion and summary, respectively.

\section{Simulation setup} \label{sec:setup}

\subsection{Test particle simulations}
\begin{figure}[h]
\centering	
\includegraphics[width=7cm]{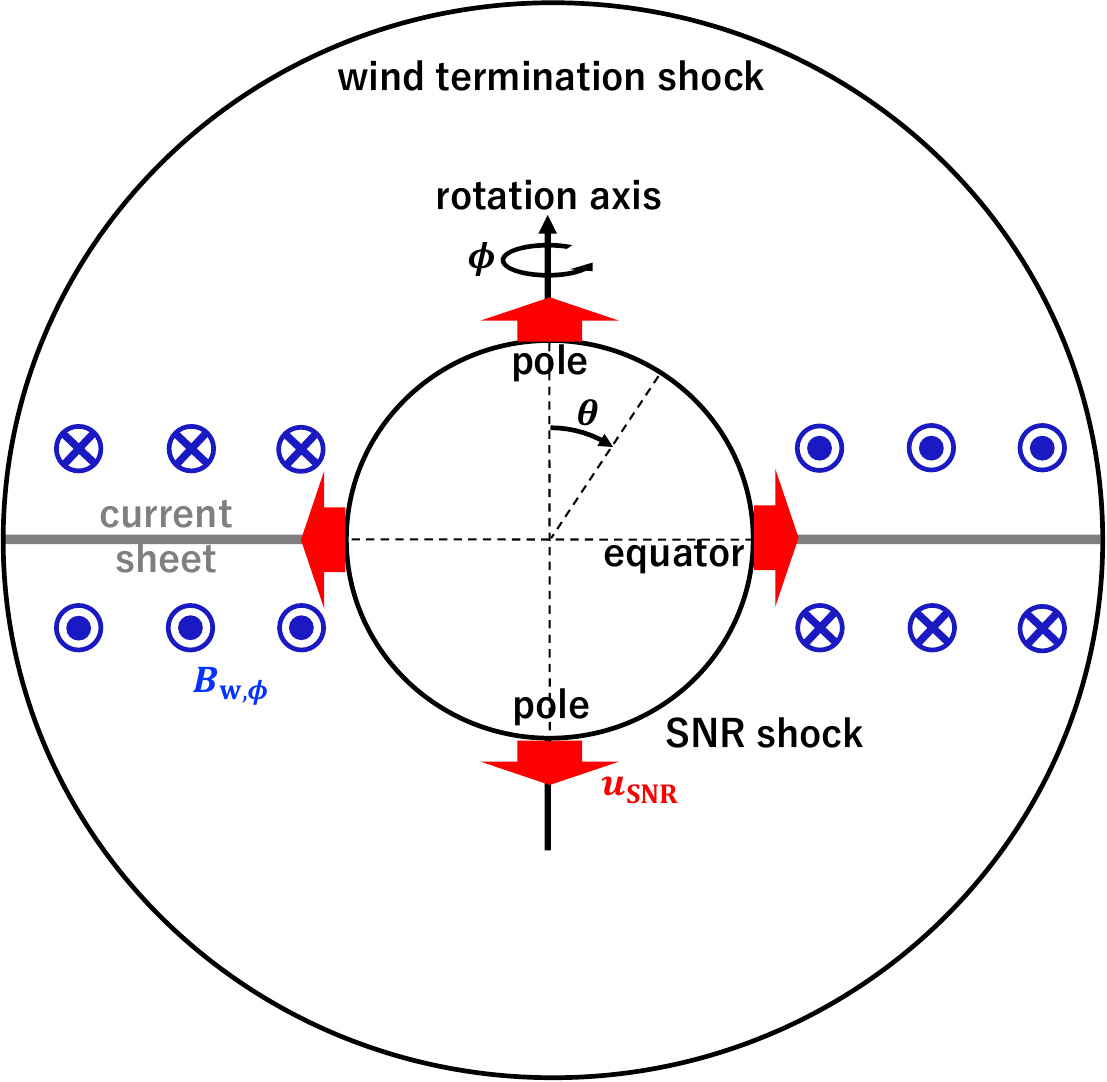}
\caption{
Schematic picture of the aligned rotator case. 
The inner and outer black circles are the SNR shock and WTS, respectively.
$\theta$ and $\phi$ are polar and azimuthal angles, respectively.
The regions at $\theta = 0, \pi$ and $\theta = \pi/2$ are the poles and equator, respectively.
The gray line at the equator is the current sheet.
$B_{{\rm w},\phi}$ is the toroidal component of the Parker-spiral magnetic field in the free wind region.
The black solid arrow is the rotation axis of progenitors.
\label{fig:aligned}}
\end{figure}

In this work, we consider propagation of a core-collapse SNR shock in the CSM with the Parker-spiral magnetic field and current sheet until the SNR shock collides with the WTS.
Figure~\ref{fig:aligned} shows the schematic picture of the case of aligned rotators.
The inner and outer black circles are the SNR shock and WTS, respectively.
The region between the SNR shock and WTS is the free wind region.
The SNR shock with the shock velocity, $u_{\rm SNR}$, expands in the free wind region until the SNR shock collides with the WTS. 
$B_{{\rm w},\phi}$ is the toroidal component of the Parker-spiral magnetic field in the free wind region.
The black solid arrow is the rotation axis of the progenitors.
$\theta$ and $\phi$ are the polar and azimuthal angles, respectively.
The regions at $\theta = 0, \pi$ and at $\theta = \pi/2$ are the poles and equator, respectively.
The gray line at the equator is the current sheet.
The quantities in the wind region are assumed to be spherically symmetric except for the Parker-spiral magnetic field in the free wind region, $\vec{B}_{\rm w}$.
We perform test particle simulations to reveal the particle acceleration and attainable maximum energy in this system. 
The numerical setups in this work are almost the same as that in our previous work \citep{kamijima22}.
The important difference of numerical setups from our previous work is that the structure of the WTS is included although our previous work did not consider the WTS because we focused on the escape process of accelerated particles from the SNR shock.
In our test particle simulations, protons with energy much larger than the thermal energy are considered. 
We focused on WR stars, which are thought to be progenitors of type Ib/Ic supernovae.
The WTS of red supergiants (RSGs) is not considered because the WTS of RSGs does not play an important role in this work (see Sec.~\ref{ssec:cond} for details).
Particles with $100~{\rm GeV}$ are uniformly injected on the whole SNR shock surface only at $100~{\rm yr}$ after the supernova explosion, $t_{\rm inj} = 100~{\rm yr}$.
To improve the statistics of high-energy particles, the particle splitting method is used in our simulations.

In this work, the magnetic field fluctuation in the free wind region (shock upstream region) is assumed to be zero.
Then, the magnetic field in the free wind region consists only of the Parker-spiral magnetic field in our simulations (see Sec.~\ref{ssec:bwind}).
In the free wind region, we solve the equation of motion in the explosion center rest frame, $d \vec{u}/dt =(e/m_{\rm p}) [ \vec{E}_{\rm w} + \vec{u}/(\gamma c) \times \vec{B}_{\rm w} ]$, to calculate the particle trajectory, where $\vec{u}$ and $\gamma = \sqrt{1 + (u/c)^2}$ are the spatial three components of four velocities and Lorentz factor of particles, respectively.
$\vec{E}_{\rm w}$ and $\vec{B}_{\rm w}$ are the electric and magnetic fields in the free wind region measured in the explosion center rest frame.
Contrary to the free wind region, the magnetic field fluctuation in the downstream region of both the SNR shock and WTS is assumed to be highly turbulent.
In the downstream region of both the SNR shock and WTS, we consider only the magnetic field strength although the magnetic field configuration is not considered.
The magnetic field strength in the downstream region of both the SNR shock and WTS is given by the condition that a fraction of the upstream kinetic energy flux is converted to the downstream magnetic field energy flux in the shock rest frame (see Sec.~\ref{ssec:bdown}).
In the downstream region of both the SNR shock and WTS, the particle trajectory is solved by the Monte-Carlo method.
The downstream particle motion is assumed to be the Bohm diffusion under the highly turbulent magnetic field and downstream particles are isotropically scattered in the downstream rest frame. 
Thus, the new momentum direction is chosen randomly, regardless of the previous direction and the downstream particles move rectilinearly between scatterings. 
The mean-free path of downstream particles is given by the downstream gyroradius because the particle motion in the downstream region is assumed to be the Bohm diffusion.

\subsection{Dynamics of the supernova remnant shock and wind termination shock}
Both the SNR shock and WTS are assumed to be spherical discontinuities.
This is because the gyroradius of high-energy protons focused in our simulations is assumed to be larger than the width of the shock.
The SNR shock velocity, $u_{\rm SNR} (t)$, is given as follows \citep{chevalier82}:
\begin{eqnarray}
u_{\rm SNR} (t) &=& 
	\left\{
	\begin{array}{ll}
	\displaystyle
	\frac{n - 3}{n - 2} \left[ \frac{2}{n(n - 4)(n - 3)} \right. & \\
	\displaystyle
	\left. \times \frac{[10(n - 5)E_{\rm SN}]^{ \frac{n - 3}{2} }}{[3(n - 3)M_{\rm ej}]^{ \frac{n - 5}{2} }} \frac{V_{\rm w}}{\dot{M}t} \right]^{\frac{1}{n - 2}} & (t \le t_{\rm t}) \\
	\displaystyle
	 \sqrt{\frac{ 2 E_{\rm SN}}{M_{\rm ej}} } \left( 1 + 2\sqrt{ \frac{2 E_{\rm SN}}{M_{\rm ej}^3} } \frac{\dot{M}}{V_{\rm w}} t \right)^{-\frac{1}{2}}  & (t \ge t_{\rm t}) 
	\end{array} 
	\right.,~~~ \label{eq:ush}
\end{eqnarray}
where $E_{\rm SN} = 10^{51}~{\rm erg}$, $M_{\rm ej} = 5M_\odot$, $\dot{M} = 10^{-4}M_\odot/{\rm yr}$, and $V_{\rm w} = 3000~{\rm km/s}$ are the explosion energy, ejecta mass,  mass loss rate, wind velocity of WR stars, respectively \citep{crowther07,niedzielski02}.
$t$ is the elapsed time from the supernova explosion.
The density in the free wind region and ejecta profile are assumed to be 
\begin{eqnarray}
\rho_{\rm w} &=& \frac{\dot{M}}{4 \pi V_{\rm w} r^2}, \label{eq:rho_w} \\
\rho_{\rm ej} &\propto& 
	\left\{
	\begin{array}{ll}
	\displaystyle
	r^0 t^{-3}& ({\rm inner~ejecta}) \\
	\displaystyle
	r^{-n} t^{n-3}& ({\rm outer~ejecta})
	\end{array}
	\right.,
\end{eqnarray}
where $n$ is set to be 10 \citep{chevalier82}.
$r$ and $t_{\rm t}$ are the distance from the explosion center and the time when the reverse shock reaches the inner ejecta:
\begin{eqnarray}
t_{\rm t} = \frac{2}{n(n-4)(n-3)} \frac{ [3(n-3)M_{\rm ej}]^{\frac{3}{2}}}{ [10(n-5)E_{\rm ej}]^{\frac{1}{2}} } \frac{V_{\rm w}}{\dot{M}}.
\end{eqnarray}
$R_{\rm SNR} = \int^t u_{\rm SNR}(t') dt'$ is the SNR shock radius.
As for the velocity profile of the downstream region of the SNR, $u_{\rm d,SNR} (r,t)$, we use the following approximate formula:
\begin{equation}
u_{\rm d,SNR} (r,t) = \frac{3 u_{\rm SNR}(t) + V_{\rm w}}{4} \left( \frac{r}{R_{\rm SNR}(t)}\right),
\end{equation}
where $u_{\rm d,SNR} (R_{\rm SNR},t) = (3 u_{\rm SNR}(t) + V_{\rm w})/4$ is derived from the Rankin-Hugoniot relation at the strong shock limit.
Particles in the downstream region of the SNR shock lose their energy by the adiabatic cooling because of the expansion of the downstream region of the SNR shock (${\rm div} \vec{u}_{\rm d,SNR}>0$).

The self-similar solution is used as the radius of the WTS \citep{weaver77}.
The radius of the WTS, $R_{\rm WTS}$, is determined by the condition that the ram pressure in the free wind region at the WTS, $\rho_{\rm w}(R_{\rm WTS}) V_{\rm w}^2$, is equal to the pressure of the shocked wind region, $P(t + t_{\rm life}) \approx P(t_{\rm life})$.
$t_{\rm life} \sim 10^5~{\rm yr}$ is the lifetime of WR stars.
$t + t_{\rm life}$ is almost the same as $t_{\rm life}$ because  $t_{\rm life}$ is much larger than the SNR age.
Hence, $R_{\rm WTS}(t_{\rm life}) = \sqrt{\dot{M}V_{\rm w}/(4\pi P(t_{\rm life}))}$ is calculated as follows \citep{weaver77,dwarkadas07}:
\begin{eqnarray}
R_{\rm WTS}(t_{\rm life}) &\approx& 0.78 \dot{M}^{3/10} V_{\rm w}^{1/10} \rho_0^{3/10} t_{\rm life}^{2/5} \label{eq;rwts}  \\
&\approx& 8~{\rm pc} \left( \frac{\dot{M}}{10^{-4}M_\odot/{\rm yr}} \right)^{3/10}  \left( \frac{V_{\rm w}}{3000~{\rm km/s}} \right)^{1/10} \nonumber \\
&& \times \left( \frac{\rho_0}{1.67\times10^{-24}~{\rm g/cm^3}} \right)^{-3/10} \left( \frac{t_{\rm life}}{10^5~{\rm yr}} \right)^{2/5}. \nonumber
\end{eqnarray}
The time evolution of the WTS can be negligible between the time of the supernova explosion and the time when the SNR collides with the WTS because the dynamical timescale is much shorter than the SNR age.
Therefore, $R_{\rm WTS}$ is fixed to be $8~{\rm pc}$ in our simulations.
The density in the shocked wind region is constant because the pressure in the shocked wind region is constant and the shocked wind region is adiabatic. 
The following velocity profile in the shocked wind region, $u_{\rm d,WTS}(r)$, is given by the mass flux conservation between the upstream and downstream regions of the WTS \citep{weaver77}:
\begin{eqnarray}
u_{\rm d,WTS}(r) = \frac{V_{\rm w}}{4} \left( \frac{r}{R_{\rm WTS}} \right)^{-2}. \label{eq:udwts}
\end{eqnarray}
$u_{\rm d,WTS}(R_{\rm WTS}) = V_{\rm w}/4$ is derived from the Rankin-Hugoniot relation at the strong shock limit.
In the shocked wind region, the adiabatic cooling does not work because ${\rm div} \vec{u}_{\rm d,WTS} = 0$.

\subsection{Magnetic field in the free wind region} \label{ssec:bwind}
The electromagnetic field in the free wind region (shock upstream region) is shown in this section.
As with our previous work \cite{kamijima22}, for simplicity, we consider only the Parker-spiral magnetic field as an unperturbed magnetic field.
Thus, we do not consider the magnetic field fluctuation in the free wind region.
The rotation axis of progenitors is set to be the polar axis of the spherical coordinate.
$\theta$ and $\phi$ are the polar and azimuthal angles.
$\phi$ is the same direction as the rotation of progenitors.
The regions at $\theta = 0,\pi$ and at $\theta = \pi/2$ are the polar and equatorial regions, respectively.
The magnetic field in the free wind region, $\vec{B}_{\rm w} = B_{{\rm w},r} \vec{e}_r + B_{{\rm w},\phi} \vec{e}_{\phi}$, is as follows \citep{weber67}:
\begin{eqnarray}
B_{{\rm w},r} &=& B_{\rm A} \left( \frac{R_{\rm A}}{r} \right)^2 \left\{ 1 - 2 H( \theta - \theta_{\rm CS} ) \right\}, \label{eq:br} \\ 
B_{{\rm w},\phi} &=& - B_{\rm A} \frac{R_{\rm A}}{r} \frac{R_{\rm A} \Omega_{*}}{V_{\rm w}} \sin \theta \left\{ 1 - 2 H( \theta - \theta_{\rm CS} ) \right\},~~~~\label{eq:bphi} 
\end{eqnarray}
where $H(\theta)$, $\Omega_* = 2\pi/P_*$, $R_{\rm A}$, and $B_{\rm A}$ are the Heaviside step function, angular frequency of progenitors, Alfv\'en radius, and the magnetic field strength at the Alfv\'en radius, respectively.
$\vec{e}_r$ and $\vec{e}_{\phi}$ are unit vectors of $r$ and $\phi$ directions.
The rotation period of progenitors, $P_*$, is set to be $10~{\rm days}$ \citep{chene08}.
$R_{\rm A}$ is approximately given by
\begin{eqnarray}
\frac{R_{\rm A}}{R_*} \approx 1 + \left( \eta_* + \frac{1}{4} \right)^\frac{1}{2q-2} - \left(\frac{1}{4} \right)^\frac{1}{2q-2}, \label{eq:ra}
\end{eqnarray}
where $R_*$ is the radius of progenitors \citep{ud-doula08}.
$\eta_* = B_*^2R_*^2/(\dot{M}V_{\rm w})$ is called as the magnetic confinement parameter \citep{ud-doula08}.
$B_*$ is the surface magnetic field strength of progenitors.
$q$ is the index of the radial dependency of the magnetic field inside the Alfv\'en radius.
In this work, we assume that the magnetic field configuration inside the Alfv\'en radius is the dipole magnetic field ($q=3$).
The Alfv\'en radius, $R_{\rm A}$, is 
\begin{eqnarray}
R_{\rm A}&=& \left\{ \begin{array}{ll}
R_* & ~(~\eta_* \ll 1~) \\
R_*\eta_*^{1/4}  & ~(~\eta_* \gg 1~) \\
\end{array} \right. \\
&=& \left\{ \begin{array}{ll}
R_* & ~(~\eta_* \ll 1~) \\
R_*^{\frac{3}{2}} B_*^{\frac{1}{2}}\dot{M}^{-\frac{1}{4}} V_{\rm w}^{-\frac{1}{4}}& ~(~\eta_* \gg 1~) \\
\end{array} \right..
\label{eq:ra}
\end{eqnarray}
The magnetic field strength at the Alfv\'en radius, $|B_{\rm A}|$, is 
\begin{eqnarray}
|B_{\rm A}|= \left\{ \begin{array}{ll}
B_* & ~(~\eta_* \ll 1~) \\
B_*^{-\frac{1}{2}} R_*^{-\frac{3}{2}} \dot{M}^{\frac{3}{4}} V_{\rm w}^{\frac{3}{4}}& ~(~\eta_* \gg 1~) \\  
\end{array} \right..
\label{eq:ba}
\end{eqnarray}
In this work, $B_*$ and $R_*$ are set to be $100~{\rm G}$ and $10R_\odot$, respectively \citep{hubrig20,crowther07,niedzielski02}.
The sign of $B_{\rm A} = \pm B_*\left( R_*/R_{\rm A}\right)^q$ is given by whether the angle between the rotation and magnetic axes, $\alpha_{\rm inc}$, is larger than $\pi/2$ radian.
When $\alpha_{\rm inc} \le(\ge) \pi/2$, the sign of $B_{\rm A}$ is positive (negative).
In this work, $\alpha_{\rm inc}$ is set to be $0, \pi/6, 5\pi/6$, and $\pi$.
Aligned rotators are the case of $\alpha_{\rm inc} = 0$ or $\pi$, and oblique rotators are the case of $\alpha_{\rm inc} \neq 0$ or $\pi$.
The position of the current sheet, $\theta_{\rm CS}$, is determined as follows \citep{alanko-huotari07}: 
\begin{eqnarray}
\theta_{\rm CS} = \frac{\pi}{2} - \sin^{-1} \left[ \sin\alpha_{\rm inc} \sin \left\{ \phi + \Omega_*\left( t - \frac{ r - R_{\rm A} }{V_{\rm w}} \right) \right\} \right].~~
\label{eq:thetacs}
\end{eqnarray}
The width of the current sheet is assumed to be infinitely thin because the gyroradius of high-energy protons is assumed to be much larger than the width of the current sheet.
The wind velocity is assumed to have only the radial component ($\vec{V}_{\rm w} = V_{\rm w}\vec{e}_r$).
Hence, in  the simulation frame (explosion center rest frame), the electric field, $\vec{E}_{\rm w} = -(\vec{V}_{\rm w}/c) \times \vec{B}_{\rm w}$, emerges.

\subsection{Downstream magnetic field strength} \label{ssec:bdown}
As we mentioned above, the magnetic field in the downstream region of both the SNR shock and WTS is assumed to be highly turbulent, which is suggested by recent observations and simulations \citep{prajapati19,berezhko03,bamp}.
The downstream magnetic field strength is determined by the condition that a fraction, $\epsilon_B$, of the upstream kinetic energy flux measured in the shock rest frame is converted to the downstream magnetic field energy flux.  
In this work, $\epsilon_B$ is set to be $0.1$ for both the SNR shock and WTS.
As for the SNR shock, the upstream kinetic energy flux measured in the SNR shock rest frame is $(1/2)\rho_{\rm w}(R_{\rm SNR}) (u_{\rm SNR} - V_{\rm w})^3$ and the energy flux of the  downstream magnetic field, $B_{\rm d,SNR}$, is $(B_{\rm d,SNR}^2/(8\pi))\times (u_{\rm SNR} - V_{\rm w})/4$.
Thus, $B_{\rm d,SNR}$ is 
\begin{eqnarray} 
	B_{\rm d,SNR} &=& \sqrt{ \frac{4\epsilon_B \dot{M}}{V_{\rm w}} } \frac{u_{\rm SNR} - V_{\rm w}}{R_{\rm SNR}}  \label{eq:b2snr} \\ 
	&\approx& 470~{\rm \mu G} \left( \frac{\epsilon_{\rm B}}{0.1} \right)^{1/2} \left( \frac{\dot{M}}{10^{-4} M_\odot/{\rm yr}} \right)^{1/2} \nonumber \\ 
	&&\times \left( \frac{V_{\rm w}}{3000~{\rm km/s}} \right)^{-1/2} \left( \frac{u_{\rm SNR} - V_{\rm w}}{5000~{\rm km/s}} \right) \nonumber \\
	&&\times \left( \frac{R_{\rm SNR}}{1~{\rm pc}} \right)^{-1}.~~~~~~~~
\end{eqnarray}
As for the WTS, the upstream kinetic energy flux measured in the WTS rest frame is $(1/2)\rho_{\rm w}(R_{\rm WTS}) V_{\rm w}^3$ and the energy flux of the downstream magnetic field, $B_{\rm d,WTS}$, is $(B_{\rm d,WTS}^2/(8\pi))\times V_{\rm w}/4$.
Here, the velocity of the WTS is negligible because the velocity of the WTS is much smaller than the wind velocity.
Then, $B_{\rm d,WTS}$ is 
\begin{eqnarray} 
	B_{\rm d,WTS} &=& \frac{\sqrt{4\epsilon_B \dot{M} V_{\rm w}} }{R_{\rm WTS}} \label{eq:b2wts} \\
	&\approx& 33~{\rm \mu G} \left( \frac{\epsilon_{\rm B}}{0.1} \right)^{1/2} \left( \frac{\dot{M}}{10^{-4} M_\odot/{\rm yr}} \right)^{1/5} \nonumber \\
	&&\times \left( \frac{V_{\rm w}}{3000~{\rm km/s}} \right)^{2/5} \left( \frac{\rho_0}{1.64\times10^{-24}~{\rm g/cm^3}} \right)^{3/10} \nonumber \\
	&&\times \left( \frac{t_{\rm life}}{10^5~{\rm yr}} \right)^{-2/5}.~~~ 
\end{eqnarray}

\subsection{Simulation parameters} \label{ssec:sim_param}
Simulation parameters are summarized in this section.
In our simulation, we use the magnetic field strength on the stellar surface, $B_* = 100~{\rm G}$, the stellar radius, $R_* = 10R_\odot$, the wind velocity, $V_{\rm w} = 3\times10^8~{\rm cm/s}$, the mass loss rate, $\dot{M} = 10^{-4} M_\odot/{\rm yr}$, the stellar rotation period, $P_* = 10~{\rm day}$, the supernova explosion energy, $E_{\rm SN} = 10^{51}~{\rm erg}$, the ejecta mass, $M_{\rm ej} = 5M_\odot$, the power law index of the ejecta density profile, $n = 10$, and the wind termination shock radius, $R_{\rm WTS} = 8~{\rm pc}$.
The initial SNR shock radius, $R_{\rm SNR}(t_{\rm inj} = 100~{\rm yr})$, and velocity, $u_{\rm SNR}(t_{\rm inj} = 100~{\rm yr})$, are $2.4\times10^{18}~{\rm cm}$ and $6.5\times10^8~{\rm cm/s}$, respectively.
The SNR shock velocity at the time when the SNR shock collides with the WTS ($t \approx 1450~{\rm yr}$), $u_{\rm SNR}(t \approx 1450~{\rm yr})$, is $4.7\times10^8~{\rm cm/s}$.
Under the above parameters, the magnetic confinement parameter, $\eta_* = B_*^2R_*^2/(\dot{M} V_{\rm w})$, is $2.6\times10^{-3}$.
Thus, the Alfv\'en radius, $R_{\rm A}$, and the magnetic field strength at the Alfv\'en radius, $B_{\rm A}$, are $R_*$ and $B_*$, which are determined by upper equations of Eqs.~(\ref{eq:ra}) and (\ref{eq:ba}).
At $t_{\rm inj} = 100~{\rm yr}$, the magnetic field strength around the equator of the SNR shock, $B_{\rm w}(r = R_{\rm SNR}(t_{\rm inj}), \theta \approx \pi/2) \approx B_{{\rm w},\phi}(r = R_{\rm SNR}(t_{\rm inj}), \theta \approx \pi/2)$, is about $0.48~{\rm \mu G}$.
The magnetic field strength around the equator of the WTS, $B_{\rm w}(r = R_{\rm WTS}, \theta \approx \pi/2) \approx B_{{\rm w},\phi}(r = R_{\rm WTS}, \theta \approx \pi/2)$, is about $0.048~{\rm \mu G}$.
The conversion fraction, $\epsilon_{\rm B}$, from the upstream kinetic energy flux measured in the shock rest frame to the downstream magnetic energy flux is set to be 0.1.
Then, the magnetic field strengths in the downstream regions of the WTS and SNR shock at $t = t_{\rm inj}$ are $33$ and $424~{\rm \mu G}$, respectively.
The diffusion coefficient of particles in the downstream region of the SNR shock and WTS is assumed to be the Bohm diffusion, where the downstream magnetic fields are given by Eqs.~(\ref{eq:b2snr}) and (\ref{eq:b2wts}).

\section{Theoretical estimate} \label{sec:theory}

\subsection{Cyclic motion between supernova remnant shock and wind termination shock} \label{ssec:cyclic}
\begin{figure}[h]
\centering	
\includegraphics[width=7cm]{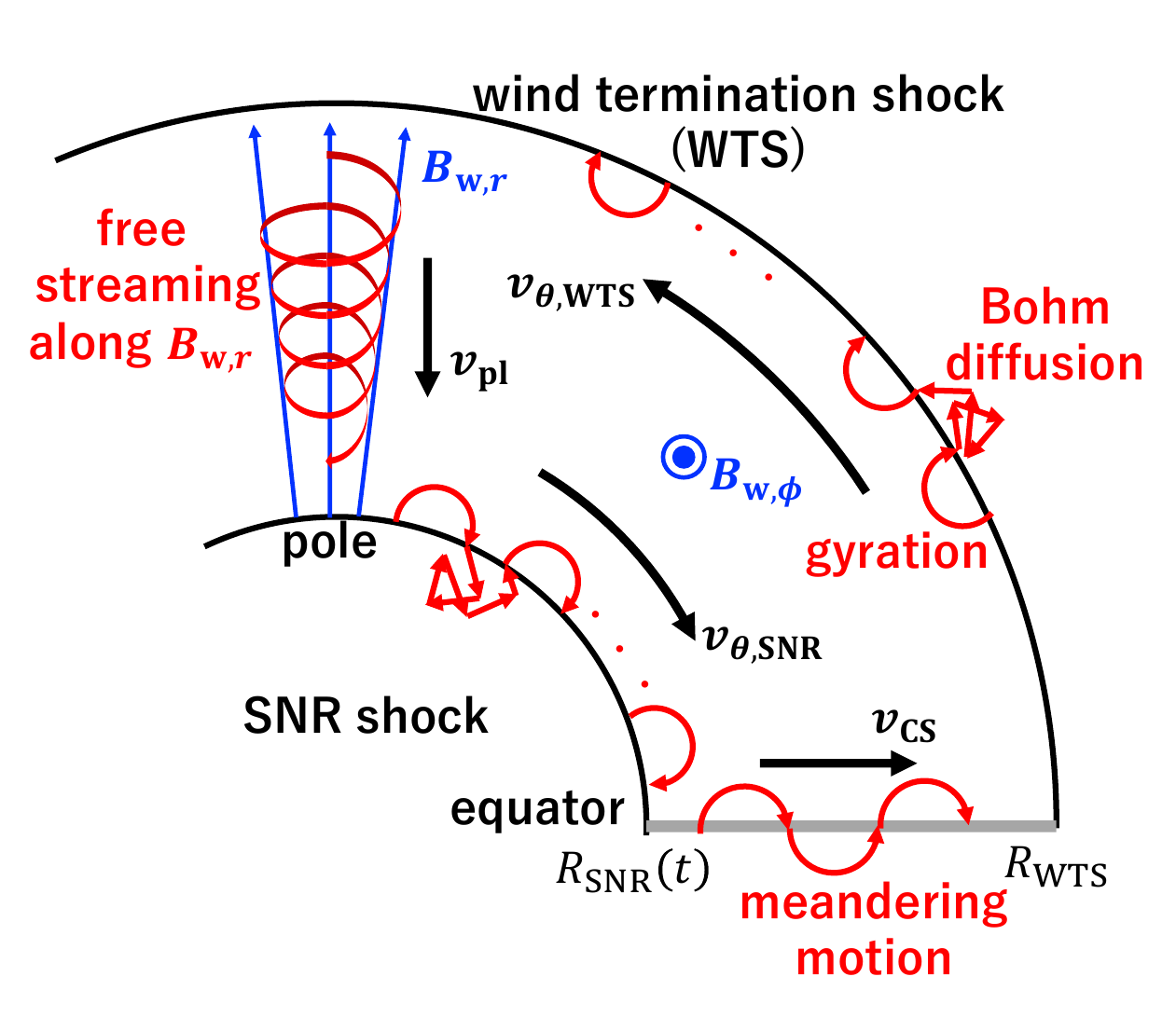}
\caption{
Schematic picture of the particle motion between the SNR shock and WTS for the aligned rotator case ($\alpha_{\rm inc} = 0$).
The red arrows show the particle orbit.
The black arrows are the mean velocities of accelerating particles along the SNR shock, $v_{\theta, \rm SNR}$, current sheet, $v_{\rm CS}$, WTS, $v_{\theta, \rm WTS}$, and pole, $v_{\rm pl}$.
$B_{{\rm w},r}$ and $B_{{\rm w},\phi}$ are the radial and toroidal components of the Parker-spiral magnetic field in the free wind region.
\label{fig:motion}}
\end{figure}
Particles with the charge, $Ze$, are considered in this section.
This section presents the global particle motion between the SNR shock and WTS, the condition for realizing this global particle motion, and the maximum attainable energy.
First, we consider the expected global particle motion between the SNR shock and WTS.
Figure~\ref{fig:motion} shows the schematic picture of the expected global particle motion between the SNR shock and WTS for the case of aligned rotators ($\alpha_{\rm inc} = 0$).
The inner and outer circles are the SNR shock and WTS, respectively.
The red arrows are the expected motion of accelerating particles.
The black arrows are the mean velocities of accelerating particles along the SNR shock, $v_{\theta, \rm SNR}$, current sheet, $v_{\rm CS}$, WTS, $v_{\theta, \rm WTS}$, and pole, $v_{\rm pl}$.
$B_{{\rm w},r}$ and $B_{{\rm w},\phi}$ are the radial and toroidal components of the Parker-spiral magnetic field in the free wind region.
When $\alpha_{\rm inc} \le (\ge) \pi/2$, particles injected on the SNR shock surface are accelerated in perpendicular shocks, and some particles are advected to the far downstream region of the SNR shock with certain probabilities.
Accelerating particles that are not advected to the far downstream region of the SNR shock move to the equator (poles) along the SNR shock.
For $\alpha_{\rm inc} \le \pi/2$, particles around the equator move to the WTS along the current sheet while performing the meandering motion \citep{levy75, kamijima22}.
For $\alpha_{\rm inc} \ge \pi/2$, particles around the poles move to the WTS along $B_{{\rm w},r}$  because $B_{{\rm w},r}$ is larger than $B_{{\rm w},\phi}$ around the poles.
Hence, for both $\alpha_{\rm inc} \le (\ge) \pi/2$, particles accelerated at the SNR shock escape from the SNR shock towards the WTS \citep{kamijima22}. 
Escaped particles eventually reach the WTS.
These particles are accelerated at the WTS and some of these particles are advected to the far downstream region of the WTS similar to the SNR shock.
Accelerating particles that are not advected to the far downstream region move to the poles (equator) along the WTS.
Here, the direction of the particle motion along the SNR shock is opposite to the direction of the particle motion along the WTS.
For $\alpha_{\rm inc} \le \pi/2$, particles that reach the poles return to the SNR shock along $B_{{\rm w},r}$.
This is because $B_{{\rm w},r}$ is larger than $B_{{\rm w},\phi}$ around the poles.
For $\alpha_{\rm inc} \ge \pi/2$, particles around the equator of the WTS are expected to return the SNR shock while moving along the current sheet (meandering motion).
Particles that return from the poles (equator) to the SNR shock can be accelerated at the SNR shock again.
These particles move to the equator (poles) and escape from the equator (poles) to the WTS again.
Thus, the cyclic particle motion between the SNR shock and WTS is expected.

We estimate the attainable energy of particles accelerated by the cyclic motion between the SNR shock and WTS, $\varepsilon_{\rm cyc}$.
$\varepsilon_{\rm cyc}$ is given by 
\begin{eqnarray}
	\varepsilon_{\rm cyc} = \sum_{\rm i=1}^{N_{\rm cyc}} ( \varepsilon_{\rm SNR}(t_{\rm i}) + \varepsilon_{\rm WTS}) \label{eq:ecyc0}.
\end{eqnarray}
$N_{\rm cyc}$ is the maximum number of cycles until the SNR shock collides with the WTS.
$\varepsilon_{\rm SNR}$ ($\varepsilon_{\rm WTS}$) is the energy gain of particles accelerated in the region between the pole and equator of the SNR shock (WTS).
$t_{\rm i}$ is the time when the {\it i} th cycle starts.

First, we estimate $\varepsilon_{\rm SNR}$.
In the SNR shock rest frame, the flow velocity in the free wind region (shock upstream region) at the time, $t$, is $u_{\rm SNR}(t) - V_{\rm w}$.
Then, the motional electric field measured in the SNR shock rest frame, $\vec{E}_{\rm w,SNR}$, emerges in the free wind region:
\begin{eqnarray}
	\vec{E}_{\rm w,SNR} &=& -\frac{u_{\rm SNR}(t) - V_{\rm w}}{c} \vec{e}_{r} \times \vec{B}_{\rm w} \nonumber \\
	&=& \frac{u_{\rm SNR}(t) - V_{\rm w}}{c} \langle B_{{\rm w},\phi} \rangle \vec{e}_{\theta}, \label{eq:ef_snr}
\end{eqnarray}
where $\vec{e}_{\theta}$ and $\langle B_{{\rm w},\phi} \rangle$ are the unit vector of the $\theta$ direction and mean magnetic field strength that particles in the free wind region feel.
For the oblique rotator case ($\alpha_{\rm inc} \neq 0,\pi$), $\langle B_{{\rm w},\phi} \rangle$ inside the wavy current sheet region ($\pi/2 - \alpha_{\rm inc} \le \theta \le \pi/2 + \alpha_{\rm inc}$) is 
\begin{eqnarray}
\langle B_{{\rm w},\phi} \rangle &\approx& - B_{\rm A} \frac{R_{\rm A}}{r} \frac{R_{\rm A} \Omega_{*}}{V_{\rm w}}  \sin \theta \left\{ 1- \frac{2}{\pi} \cos^{-1} \left( \frac{\cos \theta}{\sin \alpha_{\rm inc}} \right) \right\}, \nonumber \\
\label{eq:mean_bphi} 
\end{eqnarray}
which is derived in our previous paper \citep{kamijima22}.
Outside the wavy current sheet region ($0 \le \theta \le \pi/2 - \alpha_{\rm inc} $ and $\pi/2 + \alpha_{\rm inc} \le \theta \le \pi$), $\langle B_{{\rm w},\phi} \rangle$ is $B_{{\rm w},\phi}$ in Eq.~(\ref{eq:bphi}).
For the aligned rotator case ($\alpha_{\rm inc} = 0,\pi$), $\langle B_{{\rm w},\phi} \rangle = B_{{\rm w},\phi}$. 
In the SNR shock rest frame, particles that move along the SNR shock are accelerated by $\vec{E}_{\rm w,SNR}$ \citep{kamijima22}.
Therefore, $\varepsilon_{\rm SNR}$ is given by the potential difference between the pole and equator of the SNR shock in the SNR shock rest frame.
Thus, $\varepsilon_{\rm SNR}$ is 
\begin{eqnarray}
	\varepsilon_{\rm SNR}(t_{\rm i})	&=& \int_0^{\pi/2} Ze E_{\rm w,SNR} R_{\rm SNR} d\theta \label{eq:emax_snr_int}  \\
	&=&\left( 1 - \frac{2}{\pi} \sin \alpha_{\rm inc} \right) \frac{u_{\rm SNR}(t_{\rm i}) - V_{\rm w}}{c} \frac{Ze B_{\rm A} R_{\rm A}^2 \Omega_{*}}{V_{\rm w}}.\label{eq:emax_snr} \nonumber \\ 
\end{eqnarray}
Here, we ignored the time evolution of the SNR shock until particles injected around the poles reach the equator.
This is because the time when particles injected around the pole of the SNR shock reach the equator, $R_{\rm SNR}/v_{\theta,{\rm SNR}} \approx R_{\rm SNR}/(c/2)$, is much shorter than the dynamical timescale of the SNR shock, $R_{\rm SNR}/u_{\rm SNR}$.
$v_{\theta,{\rm SNR}}$ is the mean velocity in the $\theta$ direction of accelerating particles [see Eq.~(\ref{eq:vtheta}) in the Appendix].

Next, we estimate $\varepsilon_{\rm WTS}$.
Here, we ignored the WTS velocity because the WTS velocity is much smaller than the wind velocity.
Then, in the WTS rest frame, the motional electric field in the free wind region, $\vec{E}_{\rm w,WTS}$, is 
\begin{eqnarray}
	\vec{E}_{\rm w,WTS} &=& \frac{V_{\rm w}}{c} \langle B_{{\rm w},\phi} \rangle \vec{e}_{\theta}.~ \label{eq:ef_wts}
\end{eqnarray}
Similar to $\varepsilon_{\rm SNR}$, $\varepsilon_{\rm WTS}$ is given by the potential difference between the equator and pole of the WTS as follows:
\begin{eqnarray}
	\varepsilon_{\rm WTS} &=& \int_{\pi/2}^0 Ze E_{\rm w,WTS} R_{\rm WTS} d\theta \label{eq:emax_wts_int}  \\
	&=& \left( 1 - \frac{2}{\pi} \sin \alpha_{\rm inc} \right) \frac{Ze B_{\rm A} R_{\rm A}^2   \Omega_{*}}{c}. \label{eq:emax_wts} 
\end{eqnarray}
The maximum energies given in Eqs.~(\ref{eq:emax_snr_int}) and (\ref{eq:emax_wts_int}) correspond to the Hillas limit \citep{hillas84}, but the maximum energy given in Eq.~(\ref{eq:ecyc0}) exceeds the Hillas limit because particles move a distance longer than the size of the object.

For $\alpha_{\rm inc} \ge \pi/2$, $\varepsilon_{\rm SNR}$ and $\varepsilon_{\rm WTS}$ are given as follows:
\begin{eqnarray}
	\varepsilon_{\rm SNR}(t_{\rm i}) &=& \left(\frac{2}{\pi} \sin \alpha_{\rm inc} - 1\right) \frac{u_{\rm SNR}(t_{\rm i}) - V_{\rm w}}{c} \frac{Ze B_{\rm A} R_{\rm A}^2 \Omega_{*}}{V_{\rm w}},~~ \nonumber  \\
	\label{eq:emax_snr_pl} \\
	\varepsilon_{\rm WTS} &=& \left( \frac{2}{\pi} \sin \alpha_{\rm inc} - 1\right) \frac{Ze B_{\rm A} R_{\rm A}^2 \Omega_{*}}{c}.~~ \nonumber \\ 
	\label{eq:emax_wts_pl} 
\end{eqnarray}
$\langle B_{{\rm w},\phi} \rangle$ for $\alpha_{\rm inc} \ge \pi/2$ is the opposite sign of $\langle B_{{\rm w},\phi} \rangle$ for $\alpha_{\rm inc} \le \pi/2$ \citep{kamijima22}.

\subsection{The number of cycles and attainable energy} \label{ssec:cyclic}
The maximum number of cycles between the SNR shock and WTS, $N_{\rm cyc}$, is estimated in this section.
We consider the situation that particles injected on the SNR shock at $t = t_{\rm inj}$ continue to be accelerated until the SNR shock collides with the WTS.
Hence, we do not consider particles advected to the far downstream region of the SNR shock or WTS until the SNR collides with the WTS.
The SNR shock collides with the WTS at $t = t_{\rm col}$.
$N_{\rm cyc}$ is 
\begin{eqnarray}
	N_{\rm cyc} &=& \frac{t_{\rm col} - t_{\rm inj}}{\Delta t_{\rm cyc}},\label{eq:ncycle1} 
\end{eqnarray}
where $\Delta t_{\rm cyc}$ is one cycle time between the SNR shock and WTS (SNR $\rightarrow$ WTS $\rightarrow$ SNR).
$\Delta t_{\rm cyc}$ is given as follows:
\begin{eqnarray}
	\Delta t_{\rm cyc}(t) &=& \frac{\pi R_{\rm SNR}(t)/2}{v_{\theta, \rm SNR}} + \frac{R_{\rm WTS} - R_{\rm SNR}(t)}{v_{\rm eq}} \nonumber \\ 
	&& + \frac{\pi R_{\rm WTS}/2}{v_{\theta, \rm WTS}} + \frac{R_{\rm WTS} - R_{\rm SNR}(t)}{v_{\rm pl}} \label{eq:dtcycle0} ~~  \\
	&\approx& \left( \pi - 4 \right) \frac{R_{\rm SNR}(t)}{c} + \left( \pi + 4 \right) \frac{R_{\rm WTS}}{c} \label{eq:dtcycle1} \\ 
	&\approx& \left( \pi + 4 \right) \frac{R_{\rm WTS}}{c},\label{eq:dtcycle}
\end{eqnarray}
where $v_{\theta, \rm SNR}, v_{\rm eq}, v_{\theta, \rm WTS}$, and $v_{\rm pl}$ are the mean velocities of accelerating particles moving along the SNR shock,  equator, WTS, and poles, respectively (see Fig.~\ref{fig:motion}).
As shown in the Appendix, $v_{\theta, \rm SNR}, v_{\rm eq}, v_{\theta, \rm WTS}$, and $v_{\rm pl}$ are approximately $c/2$ if the residence time in the downstream region of the SNR shock and WTS is much smaller than that in the free wind region (shock upstream region) [see Eqs.~(\ref{eq:vtheta}),(\ref{eq:vpl}), and (\ref{eq:veq}) in the Appendix].
In our model, the downstream residence time can be negligible if the downstream magnetic field strength is much larger than the magnetic field strength in the free wind region \citep{kamijima20}.
The first term in Eq.~(\ref{eq:dtcycle0}) means the elapsed time when accelerating particles move along the SNR shock from the pole to the equator.
The second term means the elapsed time when particles move along the equator between the SNR shock and WTS.
The third term means the elapsed time when accelerating particles move along the WTS from the equator to the pole.
The fourth term means the elapsed time when particles move along the pole between the SNR shock and WTS.
We ignored the first term in Eq.~(\ref{eq:dtcycle1}) because $R_{\rm SNR} < R_{\rm WTS}$.
Therefore, $N_{\rm cyc}$ is calculated to be
\begin{eqnarray}
	N_{\rm cyc} &\approx& \frac{c \Delta t_{\rm col} }{(\pi + 4)R_{\rm WTS}} \label{eq:ncycle} \\
	 &\approx& 7.8 \left( \frac{u_{\rm SNR}}{5300~{\rm km/s}} \right)^{-1}, \label{eq:ncycle_num}
\end{eqnarray}
where $\Delta t_{\rm col} = t_{\rm col} - t_{\rm inj}$.
In this work, $t_{\rm col}$ is approximately $1500~{\rm yr}$.
Here, we assume that $t_{\rm inj}$ is much earlier than $t_{\rm col}$.
Thus, $\Delta t_{\rm col} = t_{\rm col} - t_{\rm inj} \approx t_{\rm col}$.
Furthermore, $N_{\rm cyc}$ is determined by $u_{\rm SNR}$ because $R_{\rm WTS} \approx u_{\rm SNR} t_{\rm col}$.
Then, thanks to the WTS, the attainable energy of particles accelerated in the SNR-WTS system, $\varepsilon_{\rm cyc}$, can be about 8 times the maximum energy of particles accelerated in SNR shock, $\varepsilon_{\rm SNR}$.
If particles are injected at a time close to $t_{\rm col}$, these particles cannot experience the cyclic motion until the SNR shock collides with the WTS.
$N_{\rm cyc}$ is larger than unity if $t_{\rm inj} < 1300~{\rm yr}$.
Therefore, particles injected at $t_{\rm inj} < 1300~{\rm yr}$ could experience one or more cycles between the SNR shock and WTS.
The number of cycles reduces if the downstream magnetic field amplification is not sufficient and the mean residence time in the downstream region is longer than that in the free wind region (upstream region).
The mean residence time in the free wind region (upstream region), $\left< \Delta t_{\rm w} \right> = \pi \Omega_{\rm g,w}^{-1}$, is the half of the gyroperiod, where $\Omega_{\rm g,w}$ is the gyrofrequency in the free wind region (upstream region).
The mean residence time in the downstream region, $\left< \Delta t_{\rm d} \right> = 4 \kappa_{\rm d}/(u_{\rm d}v)$, is determined by the downstream diffusion coefficient for the Bohm diffusion, $\kappa_{\rm d} = (B_{\rm w}/B_{\rm d})r_{\rm g,w}v/3$.
$r_{\rm g,w} = v\Omega_{\rm g,w}^{-1}$ and $v$ are the gyroradius in the free wind region (upstream region) and the particle velocity.
The magnetic field strengths in the downstream regions of the SNR shock, $B_{\rm d,SNR}$, and the WTS, $B_{\rm d,WTS}$, are given by Eqs.~(\ref{eq:b2snr}) and (\ref{eq:b2wts}).
The upstream magnetic field strength, $B_{\rm w} \approx B_{{\rm w},\phi}$, is given by Eq.~(\ref{eq:bphi}).
$u_{\rm d}$ is the downstream flow velocity.
The condition that the residence time in the downstream region is longer than that in the free wind region (upstream region) is 
\begin{eqnarray} 
	\frac{B_{\rm d}}{B_{\rm w}} \le \frac{16}{3\pi} \frac{c}{u_{\rm w}} \approx 170 \left( \frac{u_{\rm w}}{3000~{\rm km/s}} \right)^{-1}.
\end{eqnarray}
$u_{\rm w}$ is the upstream flow velocity measured in the SNR shock rest frame, $u_{\rm w,SNR} = u_{\rm SNR} - V_{\rm w} = 4u_{\rm d,SNR}$, or WTS rest frame, $u_{\rm w,WTS} = V_{\rm w} = 4u_{\rm d,WTS}$, where $u_{\rm SNR}$ and $V_{\rm w}$ are the SNR shock velocity measured in the explosion center rest frame [Eq.~(\ref{eq:ush})] and the wind velocity. 
From Eqs.~(\ref{eq:ecyc0}), (\ref{eq:emax_snr}), (\ref{eq:emax_wts}), and (\ref{eq:ncycle}), $\varepsilon_{\rm cyc}$ is 
\begin{eqnarray}
\varepsilon_{\rm cyc} &=&  \sum_{\rm i = 1}^{N_{\rm cyc}}  \left( \varepsilon_{\rm SNR}(t_{\rm i}) + \varepsilon_{\rm WTS} \right)~~~~~\\
	&=& \left| 1 -\frac{2}{\pi} \sin \alpha_{\rm inc} \right| \sum_{\rm i = 1}^{N_{\rm cyc}} \frac{u_{\rm SNR}(t_{\rm i})}{c} \frac{Ze \left| B_{\rm A} \right| R_{\rm A}^2 \Omega_*}{V_{\rm w}} \label{eq:emax_cycle_1} \\
	&\approx&  \left| 1 -\frac{2}{\pi} \sin \alpha_{\rm inc} \right| \frac{N_{\rm cyc} u_{\rm SNR}(t_{\rm inj})}{c} \frac{Ze \left| B_{\rm A} \right| R_{\rm A}^2\Omega_*}{V_{\rm w}}\label{eq:emax_cycle_2} \\
	&\approx& \left| 1 -\frac{2}{\pi} \sin \alpha_{\rm inc} \right| \frac{u_{\rm SNR}(t_{\rm inj}) \Delta t_{\rm col}}{(\pi + 4)R_{\rm WTS}} \frac{Ze \left| B_{\rm A} \right| R_{\rm A}^2\Omega_*}{V_{\rm w}}~~~~~~~\\
	&\approx& \left| 1 -\frac{2}{\pi} \sin \alpha_{\rm inc} \right| \frac{Ze \left| B_{\rm A} \right| R_{\rm A}^2\Omega_*}{(\pi + 4) V_{\rm w}},~~~~ \label{eq:emax_cycle}
\end{eqnarray}
where $u_{\rm SNR}(t_{\rm inj}) \Delta t_{\rm col} \approx u_{\rm SNR}(t_{\rm inj}) t_{\rm col} \approx R_{\rm WTS}$ because $t_{\rm inj} \ll t_{\rm col}$ is assumed. 
Deceleration of the SNR shock velocity can be negligible under the parameters we use in our simulations.
This is because the mass of the circumstellar matter that the SNR shock sweeps up until the SNR shock collides with the WTS is $0.1$-$1M_\odot$, which is much smaller than the ejecta mass, $M_{\rm ej} \sim 10M_\odot$.
The SNR shock velocity at the {\it i} th cycle, $u_{\rm SNR}(t_{\rm i})$, is the almost same as the SNR shock velocity at $t_{\rm inj}$, $u_{\rm SNR}(t_{\rm inj})$.
Hence, $\sum_{\rm i = 1}^{N_{\rm cyc}} u_{\rm SNR}(t_{\rm i}) \approx N_{\rm cyc} u_{\rm SNR}(t_{\rm inj})$.
As one can in see Eq.~(\ref{eq:emax_cycle}), $\varepsilon_{\rm cyc}$ depends on not SNR parameters but WR star parameters.
From Eqs.~(\ref{eq:ra}) and (\ref{eq:ba}), $\varepsilon_{\rm cyc}$ is
\begin{eqnarray}
\varepsilon_{\rm cyc} &\approx& \left| 1 -\frac{2}{\pi} \sin \alpha_{\rm inc} \right| \frac{Ze }{\pi + 4} \nonumber \\ 
&&\times \left\{ \begin{array}{ll}
B_* R_*^2 V_{\rm w}^{-1} \Omega_* & ~(~\eta_* \ll 1~) \\
B_*^{1/2} R_*^{3/2} \dot{M}^{1/4} V_{\rm w}^{-3/4} \Omega_* & ~(~\eta_* \gg 1~) \\  
\end{array} \right..~~~~~
\end{eqnarray}

\subsection{Maximum energy in the SNR-WTS system} \label{ssec:cond}
In Eq.~(\ref{eq:emax_cycle}), the attainable energy becomes large when the magnetic field strength in the free wind region of WR stars is strong ($B_{{\rm w},\phi} \propto B_{\rm A} R_{\rm A}^2\Omega_* / V_{\rm w}$).
However, the residence time in the free wind region (upstream region) becomes comparable to that in the downstream region when the magnetic field strength in the free wind region is strong.
$v_{\theta, \rm SNR}$ and $v_{\theta, \rm WTS}$ can be smaller than $c/2$ when the downstream residence time is not negligible.
This is because the particle motion in the downstream region is assumed to be diffusion and the downstream diffusion velocity is much smaller than the drift velocity in the free wind region. 
If the downstream residence time is larger than the residence time in the free wind region, particles spend most of their time in the downstream region.
Then, the main particle transport process around the shock becomes diffusion in the downstream region, so that the diffusing particles are hard to spread along the $\theta$ direction compared with the case that particles spend most of their time in the free wind region. 
Therefore, $v_{\theta, \rm SNR}$ and $v_{\theta, \rm WTS}$ can be smaller than $c/2$.
In our acceleration model, the downstream residence time can be larger than the residence time in the free wind region when the downstream magnetic field strength is not much larger than the magnetic field strength in the free wind region \citep{kamijima20}.
$v_{\theta,\rm WTS}$ rather than $v_{\theta, \rm SNR}$ can be the bottleneck for the cyclic motion between the SNR shock and WTS.
From Eq.~(\ref{eq:vtheta}), $v_{\theta,\rm WTS}$ is 
\begin{eqnarray}
	v_{\theta,\rm WTS} \approx  \frac{4c}{3\pi} \left\{ 1+ \frac{16}{3\pi} \left( \frac{ B_{\rm d,WTS}}{B_{{\rm w},\phi}(R_{\rm WTS}) } \right)^{-1} \left( \frac{V_{\rm w}}{c} \right)^{-1} \right\}^{-1}\label{eq:vtheta_wts}.~  \nonumber \\
\end{eqnarray}
When $B_{\rm d,WTS}/B_{{\rm w},\phi}(R_{\rm WTS}) \ge 16c/(3\pi V_{\rm w})$, the residence time in the downstream region of the WTS is smaller than the residence time in the free wind region.
Thus, $v_{\theta,\rm WTS}$ approximately becomes $c/2$.
This condition, $B_{\rm d,WTS}/B_{\rm w}(R_{\rm WTS}) \ge 16c/(3\pi V_{\rm w})$, can be rewritten as the condition of the rotation period of progenitors, $P_*$, as follows:
\begin{eqnarray} 
	P_* \ge P_{\rm *,cr} &=& \frac{ 16 c \left| B_{\rm A}  \right| R_{\rm A}^2 }{ 3 \epsilon_{\rm B}^{1/2} \dot{M}^{1/2} V_{\rm w}^{5/2} },~~\label{eq:cond_p}
\end{eqnarray}
where $P_{*,\rm cr}$ is the critical rotation period, which is given by the condition that the residence time in the downstream region of the WTS is equal to the residence time in the free wind region.
The shorter rotation period leads to the larger toroidal component of the Parker-spiral magnetic field in the free wind region ($B_{{\rm w},\phi} \propto P_*^{-1}$).
However, the downstream magnetic field strength in our model does not depend on $P_*$.
Then, the shorter $P_*$ leads to the smaller $B_{\rm d,WTS}/B_{{\rm w},\phi}(R_{\rm WTS})$.
The larger $B_{\rm d,WTS}/B_{{\rm w},\phi}(R_{\rm WTS})$ is favorable for the larger  $v_{\theta,\rm WTS}$, which means that the longer $P_*$ is favorable.
For the case of the longer $P_*$, the energy gain in the SNR shock becomes small due to the smaller magnetic field strength in the free wind region although $v_{\theta,\rm WTS}$ approaches $c/2$.
The attainable energy, $\varepsilon_{\rm cyc}$, becomes the maximum value, $\varepsilon_{\rm cyc,max}$, when $P_* = P_{\rm *,cr}$.
From Eqs.~(\ref{eq:emax_cycle}), (\ref{eq:cond_p}), and $P_* = 2\pi/\Omega_*$, $\varepsilon_{\rm cyc,max}$ is 
\begin{eqnarray} 
	\varepsilon_{\rm cyc,max} &\approx& \left| 1 -\frac{2}{\pi} \sin \alpha_{\rm inc} \right| \frac{3\pi Ze \epsilon_{\rm B}^{1/2} \dot{M}^{1/2} V_{\rm w}^{3/2}}{8(\pi + 4) c}~ \\
	&\approx& 216~{\rm TeV} \left| 1 -\frac{2}{\pi} \sin \alpha_{\rm inc} \right| Z \left( \frac{\epsilon_{\rm B}}{0.1} \right)^{1/2} \nonumber \\
	&&\times  \left( \frac{\dot{M}}{10^{-4}M_\odot/{\rm yr}} \right)^{1/2} \left( \frac{V_{\rm w}}{3000~{\rm km/s}} \right)^{3/2} \label{eq:emaxcyc}.~~~~~~
\end{eqnarray}
For protons ($Z  = 1$), unrealistic mass loss rate and wind velocity are required to accelerate protons to the PeV scale without the upstream magnetic field amplification.

This work focuses on the WTS of WR stars.
RSGs are also expected to create the WTS similar to WR stars.
However, the cyclic motion between the SNR shock and WTS of RSGs could be hard to occur compared with the WTS of WR stars.
This is because, in the explosion center rest frame, the velocity of the WTS of RSGs is almost the same as the wind velocity of RSGs \citep{dwarkadas05}.
The upstream kinetic energy flux measured in the WTS rest frame is much smaller than that for WR stars.
This leads to a quite small magnetic field in the downstream region of the WTS of RSGs.
Then, the residence time in the downstream region of the WTS is much larger than that in the free wind region.
Therefore, the WTS of RSGs could not play an important role in the cyclic motion between the SNR shock and WTS.

\section{Simulation results} \label{sec:result}

\subsection{Aligned rotators} \label{ssec:aligned}
\begin{figure*}[htbp]
\centering	
\includegraphics[width=15cm]{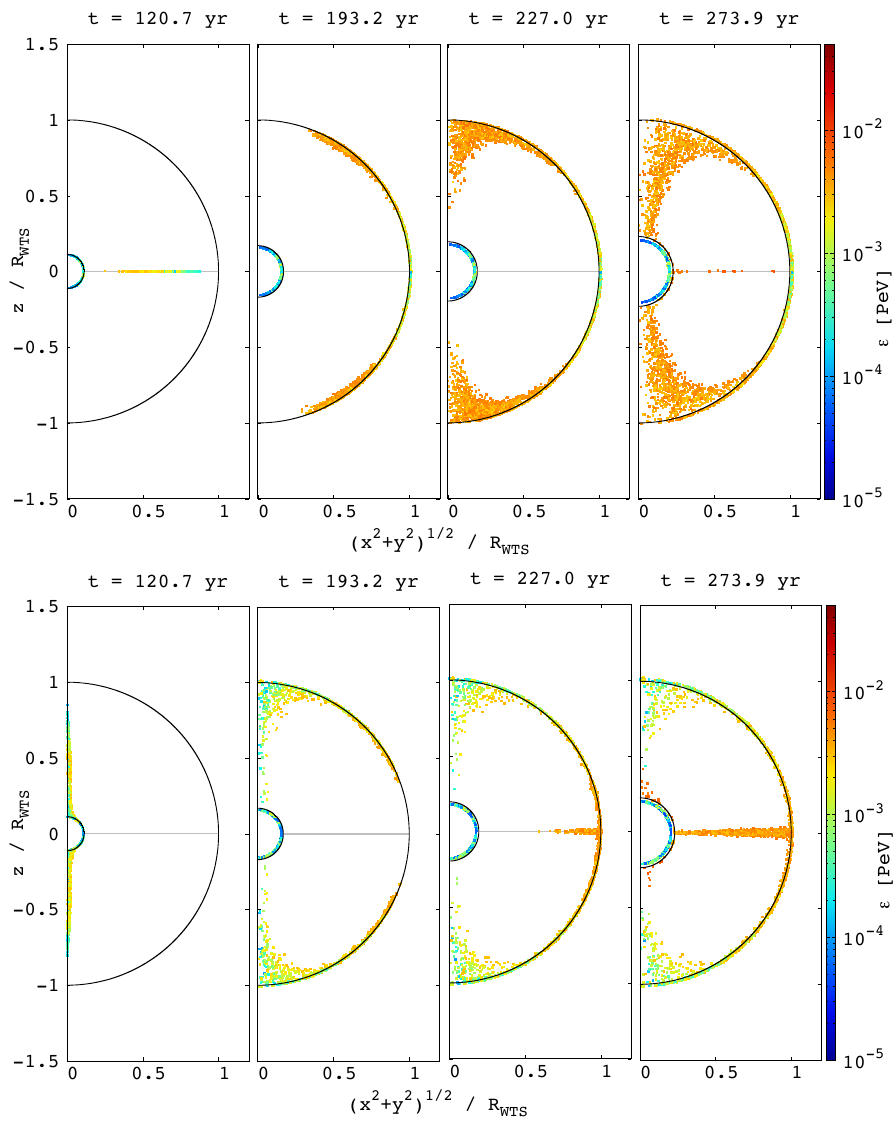}
\caption{
Time evolution of the particle distribution for aligned rotators.
The top (bottom) four panels are the results for the case of $\alpha_{\rm inc} = 0\ (\pi)$.
The vertical axis is the $z$ component of the particle position.
The $z$ direction is parallel to the rotation axis of progenitors.
The horizontal axis is the distance from the rotation axis, $\sqrt{x^2 + y^2}$.
Both axes are normalized by the radius of the WTS, $R_{\rm WTS}$.
The inner and outer black semicircles are the SNR shock and WTS, respectively.
The points and their color are particles and the particle energy, respectively.
The gray line at the equator ($z=0$) is the current sheet.
Time elapses from the left to right panels.
\label{fig:cycle_wr}}
\end{figure*}
First, we show the simulation results for the case of aligned rotators.
The top four panels in Fig.~\ref{fig:cycle_wr} are the time evolution of the particle distribution for the case of $\alpha_{\rm inc} = 0$.
The vertical axis is the $z$ component of the spatial coordinate.
The $z$ direction is parallel to the rotation axis of progenitors.
The horizontal axis shows the distance from the rotation axis of progenitors, $\sqrt{x^2 + y^2}$.
Both the horizontal and vertical axes are normalized by the WTS radius, $R_{\rm WTS}$.
The inner and outer black semicircles are the SNR shock and WTS, respectively.
The points and their color show the positions of simulation particles and their particle energies, respectively.
The gray line at the equator ($z=0$) is the current sheet.
Particles injected on the SNR shock surface are accelerated at the SNR shock while moving to the equator, reaching the equator eventually.
Once particles interact with the current sheet, particles escape from the SNR shock while moving along the current sheet (meandering motion). 
These acceleration and escape processes are shown in our previous work \citep{kamijima22}.
Particles escaped from the SNR shock move towards the WTS ($t = 120.7~{\rm yr}$ in Fig.~\ref{fig:cycle_wr}).
Particles that reach the WTS are accelerated at the WTS while moving to the poles ($t = 193.2~{\rm yr}$ in Fig.~\ref{fig:cycle_wr}).
The toroidal component of the Parker-spiral magnetic field, $B_{{\rm w},\phi}$, becomes small towards the poles.
$B_{{\rm w},\phi}$ around the poles is smaller than the radial component of the Parker-spiral magnetic field, $B_{{\rm w},r}$.
Then, the shock around the poles is the parallel shock.
Particles around the poles move along $B_{{\rm w},r}$ to the SNR shock ($t = 227.0~{\rm yr}$ in Fig.~\ref{fig:cycle_wr}).
These returned particles are accelerated at the SNR shock again while moving to the equator.
These particles eventually interact with the current sheet at the equator and escape from the SNR shock again ($t = 273.9~{\rm yr}$ in Fig.~\ref{fig:cycle_wr}).
This cycle process between the SNR shock and WTS lasts until the SNR shock collides with the WTS ($t =1449.4~{\rm yr}$) and accelerates particles between the SNR shock and WTS.
In Fig.~\ref{fig:cycle_wr}, we show the first cycle between the SNR shock and WTS.
In our simulations, the maximum number of cycles until the SNR shock collides with the WTS is about 6, which is almost consistent with the theoretical estimate of the maximum number of cycles [see Eq.~(\ref{eq:ncycle_num})].
Particles are advected and accumulated in the downstream region of the SNR shock (or WTS) although it is hard to see the accumulated downstream particles in Fig.~\ref{fig:cycle_wr} due to the short time duration.

\begin{figure*}[htbp]
\centering	
\includegraphics[width=17cm]{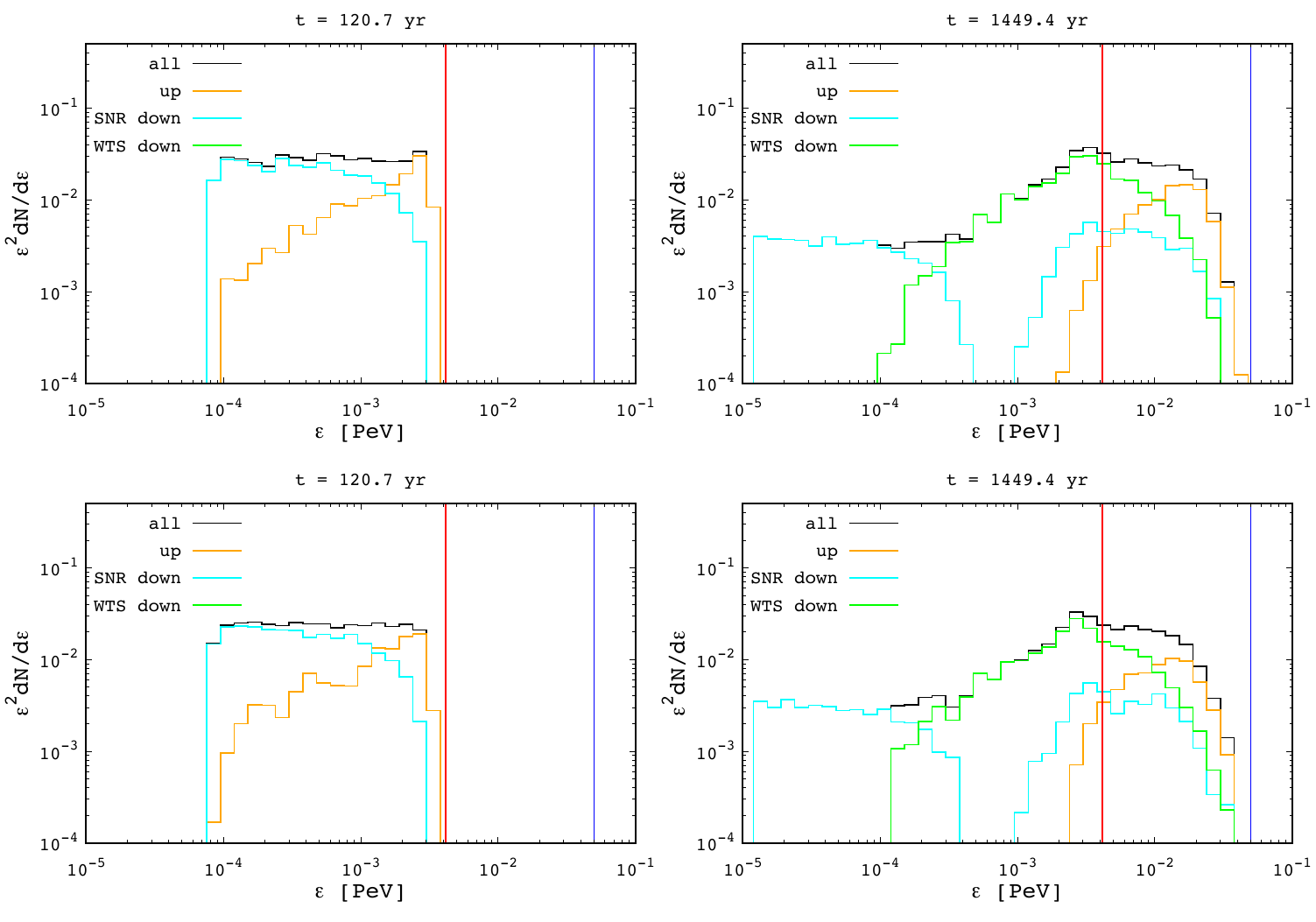}
\caption{
Energy spectra for the case of aligned rotators.
The top (bottom) two panels are results for the case of $\alpha_{\rm inc} = 0\ (\pi)$.
The black, orange, cyan, and green spectra are energy spectra of all simulation particles, simulation particles in the free wind region (upstream region), particles in the downstream region of SNR shock, and particles in the downstream region of WTS, respectively.
The horizontal and vertical axes are the particle energy, $\varepsilon$, and $\varepsilon^2dN/d\varepsilon$, respectively.
The vertical red line is the energy limited by the potential difference between the pole and equator of the SNR shock.
The vertical blue line is the energy limited by the cyclic motion between the SNR shock and WTS.
The left two panels show the energy spectra at the time when particles escaped from the SNR shock still do not reach the WTS ($t =120.7~{\rm yr}$).
The right two panels show the energy spectra at the time when the SNR shock collides with the WTS ($t =1449.4~{\rm yr}$).
\label{fig:cycle_wr_es}}
\end{figure*}
The top two panels in Fig.~\ref{fig:cycle_wr_es} show the energy spectra of all particles for the case of $\alpha_{\rm inc} = 0$.
The horizontal and vertical axes are the particle energy, $\varepsilon$, and $\varepsilon^2dN/d\varepsilon$, respectively.
The vertical red and blue lines are the energy limited by the potential difference between the pole and equator of the SNR shock (the energy gain at the SNR shock) and energy limited by the cyclic motion between the SNR shock and WTS.
In the top two panels, the values of the vertical red and blue lines are given by substituting $\alpha_{\rm inc} = 0$ into Eq.~(\ref{eq:emax_snr}) and (\ref{eq:emax_cycle}).
The top left panel shows the energy spectrum at the time when particles escaped from the SNR shock still do not reach the WTS ($t =120.7~{\rm yr}$).
The top right panel shows the energy spectrum at the time when the SNR shock collides with the WTS ($t =1449.4~{\rm yr}$).
As shown in Ref.~\cite{kamijima22}, the maximum energy of particles escaped from the SNR shock is limited by the potential difference between the pole and equator of the SNR shock ($t = 120.7~{\rm yr}$).
The reason why the cutoff energy of the energy spectrum at $t = 120.7~{\rm yr}$ is slightly smaller than the theoretical estimate of the energy gain at the SNR shock (red line) is because the number of particles injected around the poles is small due to the small solid angle around the poles although particles injected around the poles are considered in the theoretical estimate of the energy gain at the SNR shock.
Escaped particles are accelerated at the WTS while moving to the poles and the particle energy can exceed the energy gain at the SNR shock (red line).
Particles around the pole return to the SNR shock while moving along $B_{{\rm w},r}$ and are accelerated at the SNR shock again.
Thanks to this cyclic motion between the SNR shock and WTS, the maximum energy continues to increase until the SNR shock collides with the WTS ($t = 1449.4~{\rm yr}$).
The cutoff energy of the energy spectrum at $t = 1449.4~{\rm yr}$ is almost in good agreement with the theoretical estimate of the maximum energy limited by the cyclic motion (blue line) within a factor of 2. 
The energy spectrum of accelerated particles is the same as the energy spectrum of the standard DSA prediction ($dN/d\varepsilon \propto \varepsilon^{-2}$) because particles are accelerated at the SNR shock and WTS by the DSA in our simulations.
Many of the injected particles are advected to the far downstream of the SNR shock and lose their energies by the adiabatic cooling.
Thus, there are particles with energy below the injected particle energy in Fig.~\ref{fig:cycle_wr_es}.
Particles in the current sheet and poles are not magnetized and do not interact strongly with the wind flow, so that they do not suffer the adiabatic energy loss in the stellar wind although the wind is expanding (${\rm div} \vec{V}_{\rm w} \neq 0$).
As one can see in Fig.~\ref{fig:cycle_wr_es}, the energy spectra around $1~{\rm TeV}$ are harder than the standard DSA prediction.
The energy spectra around $1~{\rm TeV}$ are composed of the particles advected to the downstream region of the WTS (green spectrum at $t = 1449.4~{\rm yr}$), which originally came from the particles that escaped from the SNR shock (orange spectrum at $t = 120.7~{\rm yr}$).
The energy spectrum in the downstream region of the WTS (green spectrum) does not change due to no adiabatic energy loss in the WTS downstream region (${\rm div} \vec{u}_{\rm d,WTS} = 0$).
Contrary to the energy spectrum in the downstream region of the WTS (green spectrum), the energy spectrum in the downstream region of the SNR shock (cyan spectrum) is affected by the adiabatic energy loss because of ${\rm div} \vec{u}_{\rm d,SNR} \neq 0$.
Hence, as time goes on, the energy spectrum in the SNR shock downstream region (cyan spectrum) shifts to the low energy region although the energy spectrum in the WTS downstream region (green spectrum) does not change.
Thus, the energy spectrum harder than the standard DSA prediction emerges around $1~{\rm TeV}$.
In this work, particles are impulsively injected only at the simulation start time.
In reality, particles are continuously injected, so that the shape of the time-integrated spectrum could be different from those shown in Figs.~\ref{fig:cycle_wr_es} and \ref{fig:cycle_wr_es_30}.

Next, we show the results for the case of $\alpha_{\rm inc} = \pi$.
The bottom four panels in Fig.~\ref{fig:cycle_wr} show the time evolution of the particle distribution for the case of $\alpha_{\rm inc} = \pi$.
The bottom two panels in Fig.~\ref{fig:cycle_wr_es} show the energy spectra of all particles for the case of $\alpha_{\rm inc} = \pi$.
The sign of the magnetic field in the free wind region is opposite to that in the case of $\alpha_{\rm inc} = 0$.
Hence, the direction of the particle motion between the SNR shock and WTS is opposite to that in the case of $\alpha_{\rm inc} = 0$.
The results for $\alpha_{\rm inc} = \pi$ are the same as the results for  $\alpha_{\rm inc} = 0$ except for the direction of the particle motion.

\subsection{Oblique rotators} \label{ssec:oblique}
\begin{figure}[h]
\centering	
\includegraphics[width=7cm]{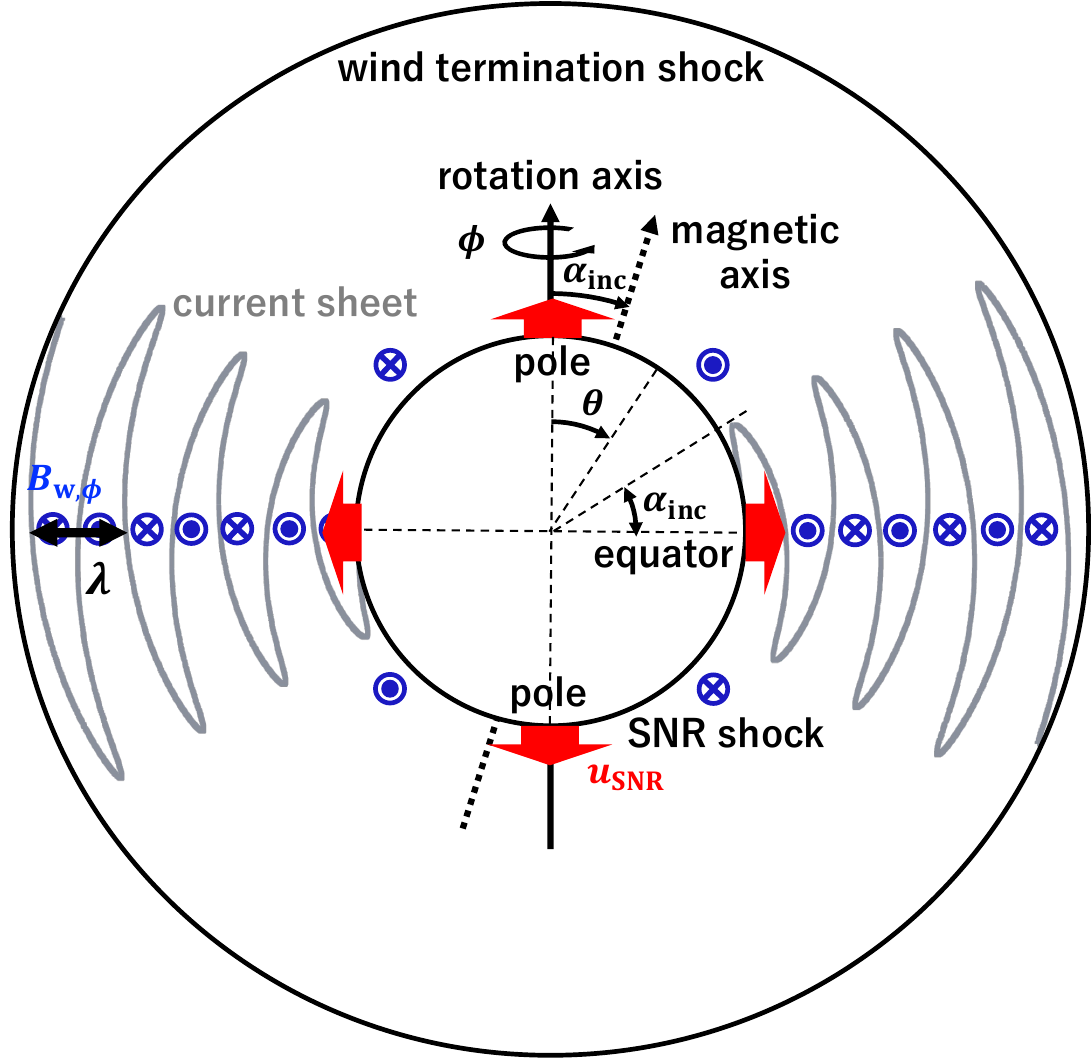}
\caption{
Schematic picture of the oblique rotator case ($\alpha_{\rm inc} \le \pi/2$). 
The inner and outer black circles are the SNR shock and WTS, respectively.
$\theta$ and $\phi$ are polar and azimuthal angles, respectively.
The regions at $\theta = 0, \pi$ and $\theta = \pi/2$ are the poles and equator, respectively.
The wavy gray line across the equator is the current sheet.
$B_{{\rm w},\phi}$ is the toroidal component of the Parker-spiral magnetic field in the free wind region.
The black solid and dotted arrows are the rotation and magnetic axes of progenitors, respectively.
$\alpha_{\rm inc}$ is the angle between the rotation and magnetic axes.
$\lambda = V_{\rm w}P_*$ is the typical length scale of the wavy current sheet.
\label{fig:oblique}}
\end{figure}
\begin{figure*}[htbp]
\centering	
\includegraphics[width=15cm]{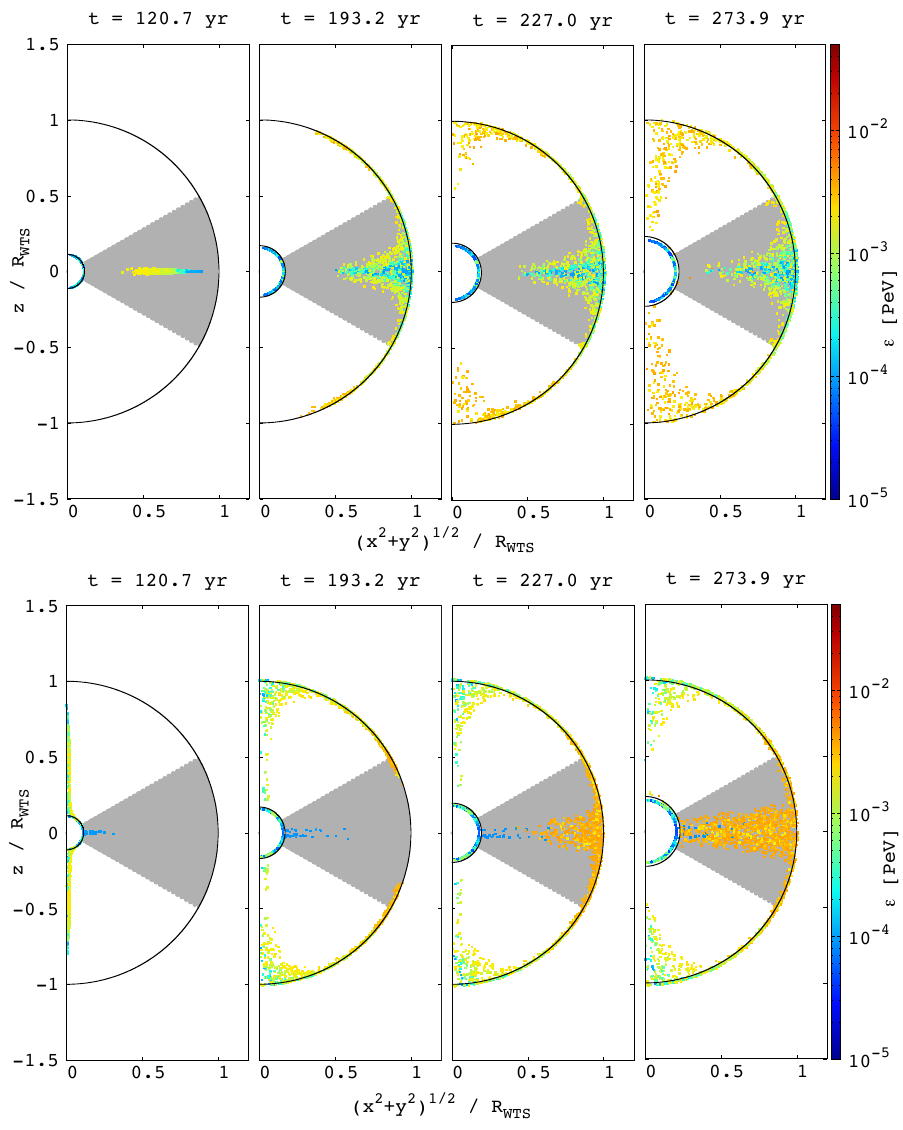}
\caption{
The same as Fig.~\ref{fig:cycle_wr}, but for $\alpha_{\rm inc} = \pi/6$ (top four panels) and $\alpha_{\rm inc} = 5\pi/6$ (bottom four panels).
The gray region across the equator ($z=0$) is the wavy current sheet region.
\label{fig:cycle_wr_30}}
\end{figure*}
The schematic picture of the oblique rotator case ($\alpha_{\rm inc} \le \pi/2$) is shown in Fig.~\ref{fig:oblique}.
The differences from aligned rotators are that the magnetic axis of progenitors is tilted by the angle, $\alpha_{\rm inc}$, from the rotation axis of progenitors and the current sheet has a wavy structure.

Next, we show the results for $\alpha_{\rm inc} = \pi/6$.
The top four panels in Fig.~\ref{fig:cycle_wr_30} are the time evolution of the particle distribution for $\alpha_{\rm inc} = \pi/6$.
The format of Fig.~\ref{fig:cycle_wr_30} is the same as that of Fig.~\ref{fig:cycle_wr}.
The gray region ($\pi/3 \le \theta \le 2\pi/3$) is the wavy current sheet region.
Particles accelerated at the SNR shock escape from the SNR shock ($t = 120.7~{\rm yr}$ in Fig.~\ref{fig:cycle_wr_30}).
Particles can escape from the SNR shock even though the current sheet is wavy.
Almost similar to the aligned rotator case, the cyclic motion between the SNR shock and WTS occurs.
However, in contract to the aligned rotator case, particles feel the magnetic field averaged over the reversal magnetic field structure inside the wavy current sheet region, which is smaller than the magnetic field for the aligned rotator case [see Eq.~(\ref{eq:mean_bphi}), and Eq.~(17) and Fig.~(10) in Ref.~\citep{kamijima22}].

\begin{figure*}[htbp]
\centering	
\includegraphics[width=17cm]{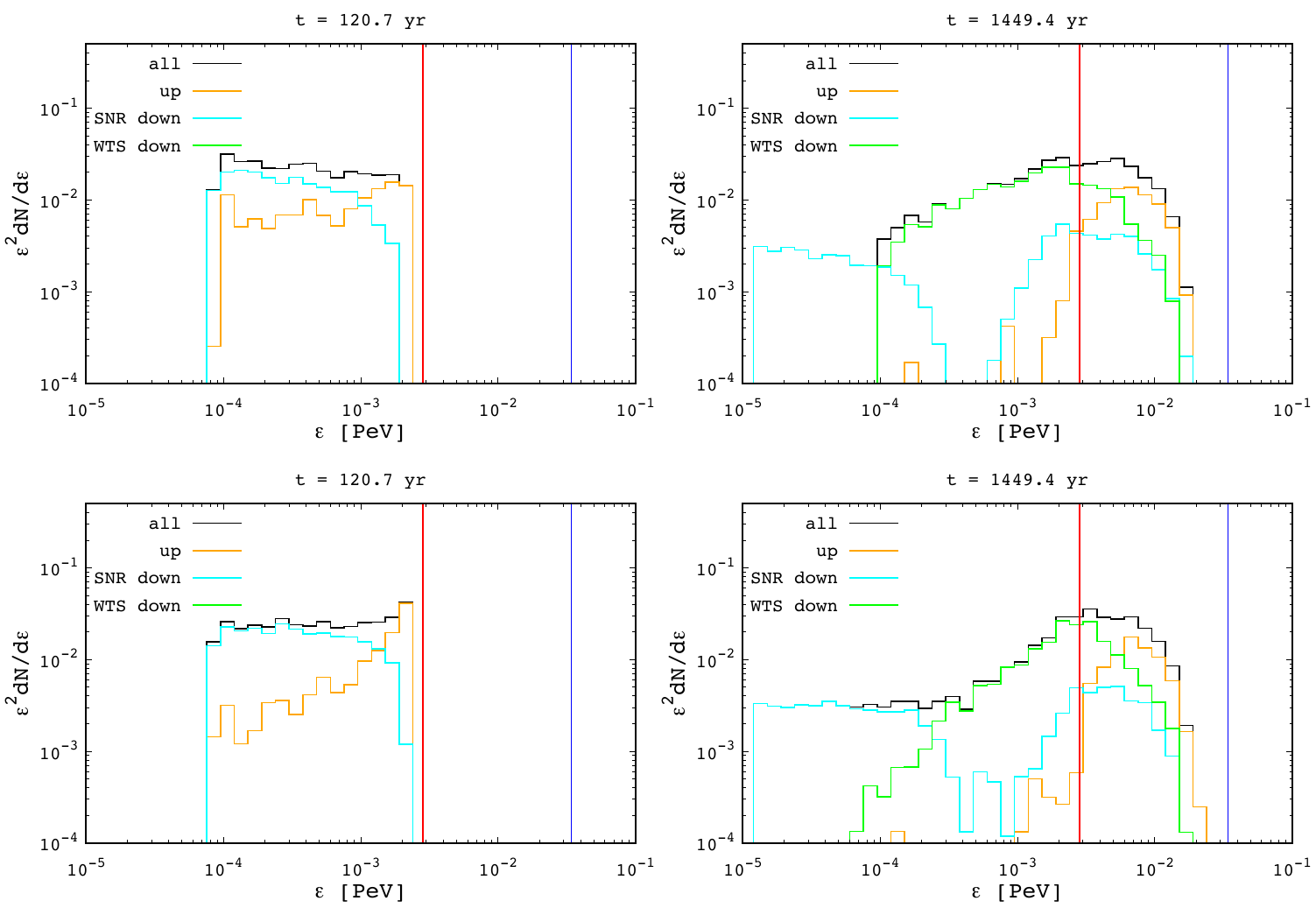}
\caption{
The same as Fig.~\ref{fig:cycle_wr_es}, but for $\alpha_{\rm inc} = \pi/6$ (top two panels) and $\alpha_{\rm inc} = 5\pi/6$ (bottom two panels).
\label{fig:cycle_wr_es_30}}
\end{figure*}
The top two panels in Fig.~\ref{fig:cycle_wr_es_30} show the energy spectra of all particles for $\alpha_{\rm inc} = \pi/6$.
Particles inside the wavy current sheet region can spread around the equator due to the weak mean magnetic field, which prevents the ideal cyclic motion.
Therefore, the deviation between the theoretical estimate of the cycle-limited maximum energy (blue line) and the cutoff energy of the energy spectrum for $\alpha_{\rm inc} = \pi/6$ is slightly larger than that for $\alpha_{\rm inc} = 0$.
This deviation at $t = 1449.4~{\rm yr}$ is within a factor of 3.
As with the case of $\alpha_{\rm inc}=0$, the energy spectrum of particles accelerated by the cyclic motion is the same as the standard DSA prediction ($dN/d\varepsilon \propto \varepsilon^{-2}$) because particles are accelerated at both the SNR shock and WTS by the DSA.

Next, we show the results for the case of $\alpha_{\rm inc} = 5\pi/6$.
The bottom four panels in Fig.~\ref{fig:cycle_wr_30} show the time evolution of the particle distribution for the case of $\alpha_{\rm inc} = 5\pi/6$.
The bottom two panels in Fig.~\ref{fig:cycle_wr_es_30} show the energy spectra of all particles for the case of $\alpha_{\rm inc} = 5\pi/6$.
The sign of the magnetic field in the free wind region is opposite to that in the case of $\alpha_{\rm inc} = \pi/6$.
Hence, the direction of the particle motion between the SNR shock and WTS is opposite to that in the case of $\alpha_{\rm inc} = \pi/6$.
The results for $\alpha_{\rm inc} = 5\pi/6$ are the same as the results for  $\alpha_{\rm inc} = \pi/6$ except for the direction of the particle motion.

\section{Discussion} \label{sec:discuss}
This work showed that the attainable energy in the SNR-WTS system could be $(c/u_{\rm SNR})/(\pi + 4)$ times the energy gain in the SNR shock.
From Eq.~(\ref{eq:emaxcyc}), a large mass loss rate and wind velocity are required to accelerate particles to the PeV scale by the cyclic motion between the SNR shock and WTS.
As we mentioned in Sec.~\ref{sec:intro}, the direct and indirect observations report the spectral break around $10~{\rm TeV}$ in the energy spectrum of observed CR protons and helium.
However, the origin of the energy scale of $10~{\rm TeV}$ is still unclear.
We investigated the escape process from type Ia SNRs and core-collapse SNRs without upstream magnetic field amplification \citep{kamijima21,kamijima22}.
For both type Ia and core-collapse SNRs without upstream magnetic field amplification, we showed that the maximum energy is limited by escape from the SNR shock and the escape-limited maximum energy is about $10~{\rm TeV}$ \citep{kamijima21,kamijima22}.
As one can see in Figs.~\ref{fig:cycle_wr_es} and \ref{fig:cycle_wr_es_30}, furthermore, particles can be accelerated to $10~{\rm TeV}$ by the cyclic motion between the SNR shock and WTS.
Thus, SNRs that upstream magnetic field amplification does not work could be the origin of the spectral break around $10~{\rm TeV}$.

The SNR shock propagates in the shocked wind region after the SNR interacts with the WTS.
The toroidal magnetic field in the shocked wind region becomes strong towards the outside of the shocked wind region until the magnetic pressure is equal to the gas pressure in the shocked wind region (Cranfil effect) \citep{cranfil}.
It is suggested that particles are accelerated to the PeV scale by the SNR shock propagating in the shocked wind region \citep{zirakashvili18}.
However, the Mach number is small because the temperature in the shocked wind region is high and the sound velocity becomes fast.
It is still unclear that particles accelerated in the shocked wind region can contribute to the observed PeV CRs because the energy spectrum of particles accelerated by the DSA in the low Mach number shock becomes softer than that for a high Mach number shock.

Particles are usually scattered in the free wind region (upstream region) due to the existence of the magnetic field fluctuation in the free wind region.
We assumed that particles are not scattered in the free wind region.
As for the acceleration process at the SNR shock and WTS, the assumption of no scattering in the free wind may not influence as long as $\delta B_{\rm w}/B_{\rm w} < 1$, where $\delta B_{\rm w}$ and $B_{\rm w}$ are the strength of the magnetic field fluctuation and the unperturbed magnetic field in the free wind region.
This is because the mean-free time in the free wind region is longer than the gyroperiod (the mean upstream residence time in our simulations) when $\delta B_{\rm w}/B_{\rm w} < 1$.
Hence, as long as $\delta B_{\rm w}/B_{\rm w} < 1$, particles are not scattered during the gyration in the free wind region and our acceleration process does not change.
On the other hand, as for the escape process from the SNR shock and the cyclic motion, the assumption of no scattering in the free wind may influence even if $\delta B_{\rm w}/B_{\rm w} < 1$.
Particles are scattered in the free wind region due to the magnetic field fluctuation.
Then, scatterings in the free wind region prevent particles escaping from the SNR shock and performing the cyclic motion between the SNR shock and WTS.

If the downstream magnetic field amplification is not sufficient and the downstream residence time is longer than the upstream residence time, the energy spectrum of accelerated particles and cyclic motion can be changed.
The anisotropic scattering in the downstream region leads to the steeper energy spectrum than the standard DSA prediction \citep{takamoto15} because most downstream particles are advected to the far downstream region while being trapped by the downstream unperturbed magnetic field.
Furthermore, if we apply the other downstream scattering model, not only the spectral index but also the cutoff shape of the energy spectrum can be affected because the spectrum cutoff shape depends on the momentum dependence of the diffusion coefficient \citep{yamazaki15}.
As we mentioned in Sec.~\ref{ssec:cyclic}, the number of cycles between the SNR shock and WTS reduces because the main particle transport process around the shock becomes diffusion in the downstream region and downstream diffusing particles are hard to spread along the $\theta$ direction compared to the case that particles spend most of their time in the free wind region.

\section{Summary} \label{sec:summary}
In this work, we have studied the particle motion between the SNR shock and WTS, and the attainable energy by using test particle simulations until the core-collapse SNR shock collides with the WTS.
The Parker-spiral magnetic field and current sheet are considered in the free wind region (shock upstream region).
We do not consider the magnetic field fluctuation and any magnetic field amplification in the free wind region.
On the other hand, the highly turbulent magnetic field is assumed in the downstream region of both the SNR shock and WTS.
We focused on WR stars as progenitors.
As shown in Ref.~\citep{kamijima22}, particles accelerated at the SNR shock escape from the equator or poles of the SNR shock towards the shock upstream region (free wind region).
The maximum energy of particles escaped from the SNR shock is limited by the potential difference between the pole and equator \citep{kamijima22}.
Escaped particles reach the WTS and move along the WTS.
For the case where the angle between the rotation and magnetic axes, $\alpha_{\rm inc}$, is smaller (larger) than $\pi/2$, we showed that particles escaped from the equator (poles) of the SNR shock are accelerated while moving to the poles (equator) of the WTS and return to the SNR shock from the poles (equator) of the WTS while moving along the radial magnetic field (current sheet).
These returned particles are accelerated at the SNR shock again.
The attainable energy given by the cyclic motion between the SNR shock and WTS is analytically derived.
The results of test particle simulations are almost in good agreement with the theoretical estimate.
The maximum energy in the SNR-WTS system is about $10$-$100~{\rm TeV}$.
Therefore, core-collapse SNRs without any magnetic field amplification in the free wind region could be the origin of the CRs that form the observed spectral break at 10 TeV.

\begin{acknowledgments}
We thank M. Hoshino, T. Amano, K. Asano, S. Imada, S. Matsukiyo, and I. Shinohara for valuable comments. 
Numerical computations were carried out on Cray XC50 at Center for Computational Astrophysics, National Astronomical Observatory of Japan. 
S.K. is supported by JSPS KAKENHI Grant No. JP22H00130. 
Y.O. is supported by JSPS KAKENHI Grants No. JP19H01893, No. JP21H04487, and No. JP24H01805.
\end{acknowledgments}

\appendix

\section{MEAN PARTICLE VELOCITY ALONG SHOCKS, POLE, AND CURRENT SHEET} \label{sec:mean_vel}
\begin{figure}[h]
\centering	
\includegraphics[width=7cm]{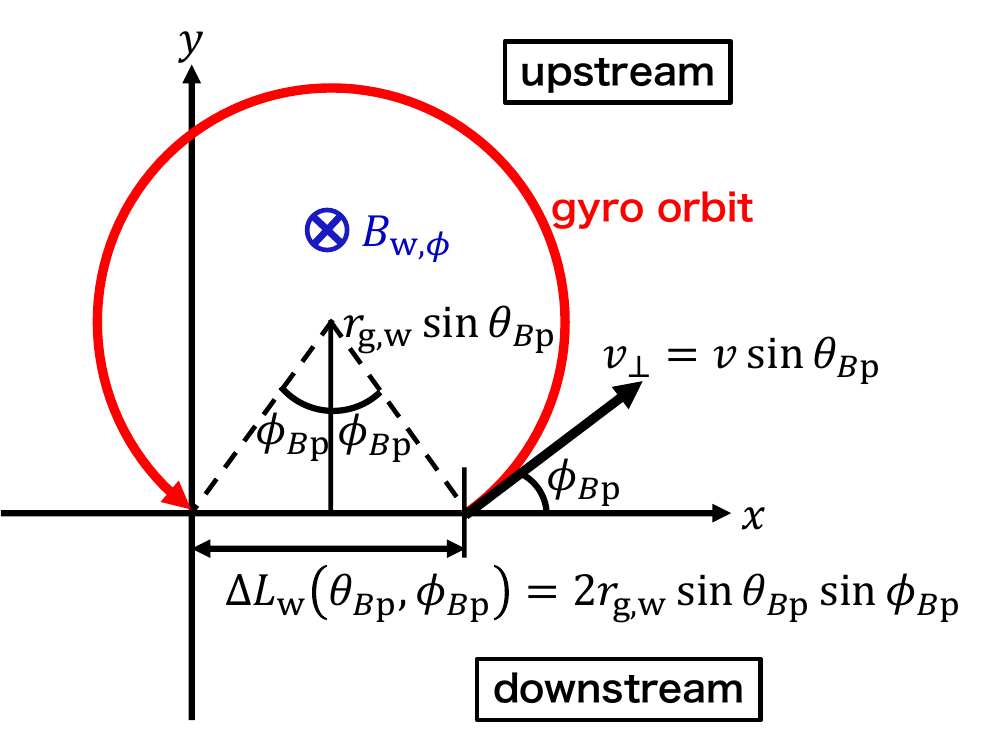}
\caption{
Particle orbit in the shock upstream region. 
The $x$ axis is the shock surface.
The positive and negative $y$ regions are the free wind region (shock upstream region) and shock downstream regions, respectively.
The red arrow is an orbit of a gyrating particle.
$\theta_{\rm Bp}$ and $\phi_{\rm Bp}$ are the pitch angle of the particle and the azimuthal angle measured from the $x$ axis, respectively.
The blue cross means the toroidal magnetic field in the free wind region, $B_{{\rm w},\phi}$.
The black arrow is the particle velocity perpendicular to $B_{{\rm w},\phi}$, $v_\perp$.
$\Delta L_{\rm w}$ is the displacement in the free wind region during the one cycle time between the free wind region (upstream region) and downstream regions, $\Delta t_{\rm w} + \Delta t_{\rm d}$.
\label{fig:dL_1}}
\end{figure}
First, we estimate the mean drift velocity of accelerating particles drifting on the shock surface.
In the explosion center rest frame (simulation frame), there is the convective electric field in the free wind region (upstream region), $\vec{E}_{\rm w} = -(\vec{V}_{\rm w}/c) \times \vec{B}_{\rm w}$. 
$\vec{E}_{\rm w}$ and $\vec{B}_{\rm w}$ are the electric and magnetic fields in the free wind region, respectively.
Particles in the free wind region experience the $\vec{E} \times \vec{B}$ drift.
However, the velocity of the $\vec{E} \times \vec{B}$ drift is $v_{E \times B} = c |\vec{E}_{\rm w} \times \vec{B}_{\rm w}|/B_{\rm w}^2 \approx V_{\rm w}$, which is much smaller than the velocities of accelerating particles ($\approx c/2$).
Thus, we ignore the electric field in the free wind region [see Eqs.~(\ref{eq:vtheta}), (\ref{eq:vpl}), and (\ref{eq:veq})]. 
In the free wind region, the accelerating particles drift on the shock surface in the $\theta$ direction towards the equator or poles depending on the magnetic field direction. 
On the other hand, the mean displacement of accelerating particles in the downstream region is zero because downstream particles are isotropically scattered in the downstream region.
Therefore, the mean displacement in the $\theta$ direction during the one back-and-forth motion is the mean displacement of particles in the free wind region, $\langle \Delta L_{\rm w} \rangle$.
The orbit of accelerating particles in the free wind region is shown in Fig.~\ref{fig:dL_1}.
The $x$ axis is the shock surface.
The positive and negative $y$ regions are the free wind region (shock upstream region) and shock downstream regions, respectively.
The red arrow is an orbit of a gyrating particle.
$\theta_{\rm Bp}$ and $\phi_{\rm Bp}$ are the pitch angle of the particle and the azimuthal angle measured from the $x$ axis, respectively.
From Fig.~\ref{fig:dL_1}, $\Delta L_{\rm w} (\theta_{B\rm p},\phi_{B\rm p})$ is equal to $ 2r_{\rm g,w} \sin \theta_{B\rm p} \sin \phi_{B\rm p}$.
$r_{\rm g,w} = \varepsilon/(Ze B_{{\rm w},\phi})$ is the gyroradius in the free wind region.
$\varepsilon$ and $B_{{\rm w},\phi}$ are the particle energy and toroidal magnetic field in the free wind region.
$\langle \Delta L_{\rm w} \rangle$ is
\begin{eqnarray}
	\langle \Delta L_{\rm w} \rangle &=& \frac{\int_0^\pi d\phi_{B\rm p} \int_0^\pi d\theta_{B\rm p}  \sin \theta_{B\rm p}  \Delta L_{\rm w} f(\theta_{B\rm p} , \phi_{B\rm p} )}{\int_0^\pi d\phi_{B\rm p}  \int_0^\pi d\theta_{B\rm p}  \sin \theta_{B\rm p}  f(\theta_{B\rm p} , \phi_{B\rm p} )} \nonumber \\
	&=& \frac{4}{3} r_{\rm g,w}, \label{eq:L1}
\end{eqnarray}
where $f(\theta_{B\rm p}, \phi_{B\rm p})$ is the particle distribution and proportional to the flux of particles that cross the shock from the downstream to the upstream, ($f(\theta_{B\rm p}, \phi_{B\rm p}) \propto v \sin \theta_{B\rm p} \sin \phi_{B\rm p}$).
The mean residence time in the free wind region is the half of the upstream gyroperiod, $\langle \Delta t_{\rm w} \rangle = \pi \Omega_{\rm g,w}^{-1} = \pi \varepsilon/(Ze B_{{\rm w},\phi} v)$.
The mean residence time in the downstream region is the same as that of the standard DSA, $\langle \Delta t_{\rm d} \rangle = 4\kappa_{\rm d}/(u_{\rm d}v)$  \citep{cr,drury83}.
$\kappa_{\rm d} = r_{\rm g,d}v/3 = (B_{{\rm w},\phi}/B_{\rm d}) r_{\rm g,w}v/3$ is the downstream diffusion coefficient, where the Bohm diffusion is assumed in the downstream region.
The downstream flow velocity in the shock rest frame, $u_{\rm d}$, is given by $(u_{\rm SNR} - V_{\rm w}) / 4$ and  $V_{\rm w}/4$ for the SNR shock and WTS, respectively, where the strong shock limit is used. 
Accelerating particles move in the $\theta$ direction by a distance of $\langle \Delta L_{\rm w} \rangle$ during $\langle \Delta t_{\rm w} \rangle + \langle \Delta t_{\rm d} \rangle$.
Hence, the mean velocity of accelerating particles in the $\theta$ direction, $v_\theta$, is 
\begin{eqnarray}
	v_\theta &=& \frac{\langle \Delta L_{\rm w} \rangle}{\langle \Delta t_{\rm w} \rangle + \langle \Delta t_{\rm d} \rangle} \nonumber \\ 
	&=& \frac{4v}{3\pi} \left\{ 1+ \frac{16}{3\pi} \left( \frac{B_{\rm d}}{B_{{\rm w},\phi}} \right)^{-1} \left( \frac{u_{\rm w}}{v} \right)^{-1} \right\}^{-1}, \label{eq:vtheta}
\end{eqnarray}
where the particle velocity, $v$, is almost the same as the speed of light, $c$, because relativistic particles are considered.
$u_{\rm w}$ is $u_{\rm SNR} - V_{\rm w}$ for the SNR shock and $V_{\rm w}$ for the WTS.
Then, $v_\theta$ is approximately $c/2$ for the case that the downstream residence time is much shorter than the residence time in the free wind region.

\begin{figure}[h]
\centering	
\includegraphics[width=7cm]{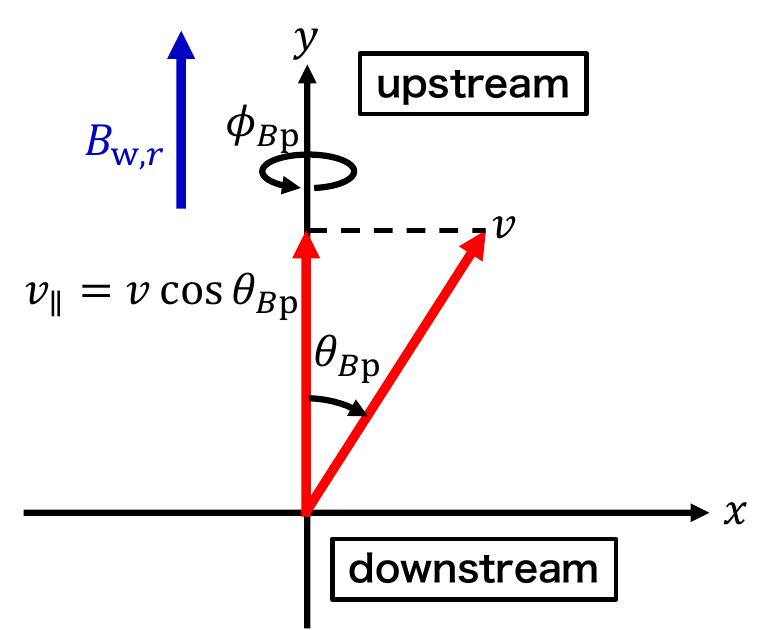}
\caption{
Geometric relationship for the particle velocity vector.
The $x$ axis is the shock surface.
The positive and negative $y$ regions are the free wind region (upstream region) and downstream region, respectively.
The red arrows are the particle velocities.
$\theta_{\rm Bp}$ and $\phi_{\rm Bp}$ are the pitch angle of a particle and azimuthal angle, respectively.
The blue arrow is the radial magnetic field region around a pole in the free wind, $B_{{\rm w},r}$, which is almost parallel to the $y$ axis.
$v_\parallel = v \cos \theta_{B\rm p}$ is the particle velocity parallel to $B_{{\rm w},r}$.
\label{fig:vpl}}
\end{figure}
Next, we estimate the mean velocity of particles that move along the radial magnetic field around the poles, $v_{\rm pl}$.
Here, we consider particles that cross the shock front from the downstream region to the free wind region (upstream region).
The geometric relationship for the particle velocity vector is shown in Fig.~\ref{fig:vpl}.
The $x$ axis is the shock surface.
The positive and negative $y$ regions are the free wind region (upstream region) and downstream region, respectively.
$\theta_{\rm Bp}$ and $\phi_{\rm Bp}$ are the pitch angle of a particle and azimuthal angle, respectively.
The blue arrow is the radial magnetic field around a pole in the free wind, $B_{{\rm w},r}$.
$v_\parallel = v \cos \theta_{B\rm p}$ is the particle velocity parallel to $B_{{\rm w},r}$.
The mean velocity parallel to $B_{{\rm w},r}$ of particles that cross the shock from the downstream region to the free wind region, $v_{\rm pl}$, is 
\begin{eqnarray}
	v_{\rm pl} &=& \frac{\int_0^{2\pi} d\phi_{B\rm p} \int_0^{\pi/2} d\theta_{B\rm p} \sin \theta_{B\rm p} v_\parallel f(\theta_{B\rm p}, \phi_{B\rm p})}{\int_0^{2\pi} d\phi_{B\rm p} \int_0^{\pi/2} d\theta_{B\rm p} \sin \theta_{B\rm p} f(\theta_{B\rm p}, \phi_{B\rm p})}~~~ \nonumber \\ 
	&=& \frac{2}{3} v,~~~ \label{eq:vpl}
\end{eqnarray}
where the particle distribution is proportional to the flux of particles that cross the shock from the downstream region to the free wind region (upstream region), $f(\theta_{B\rm p}, \phi_{B\rm p}) \propto v \cos \theta_{B\rm p}$.

\begin{figure}[h]
\centering	
\includegraphics[width=7cm]{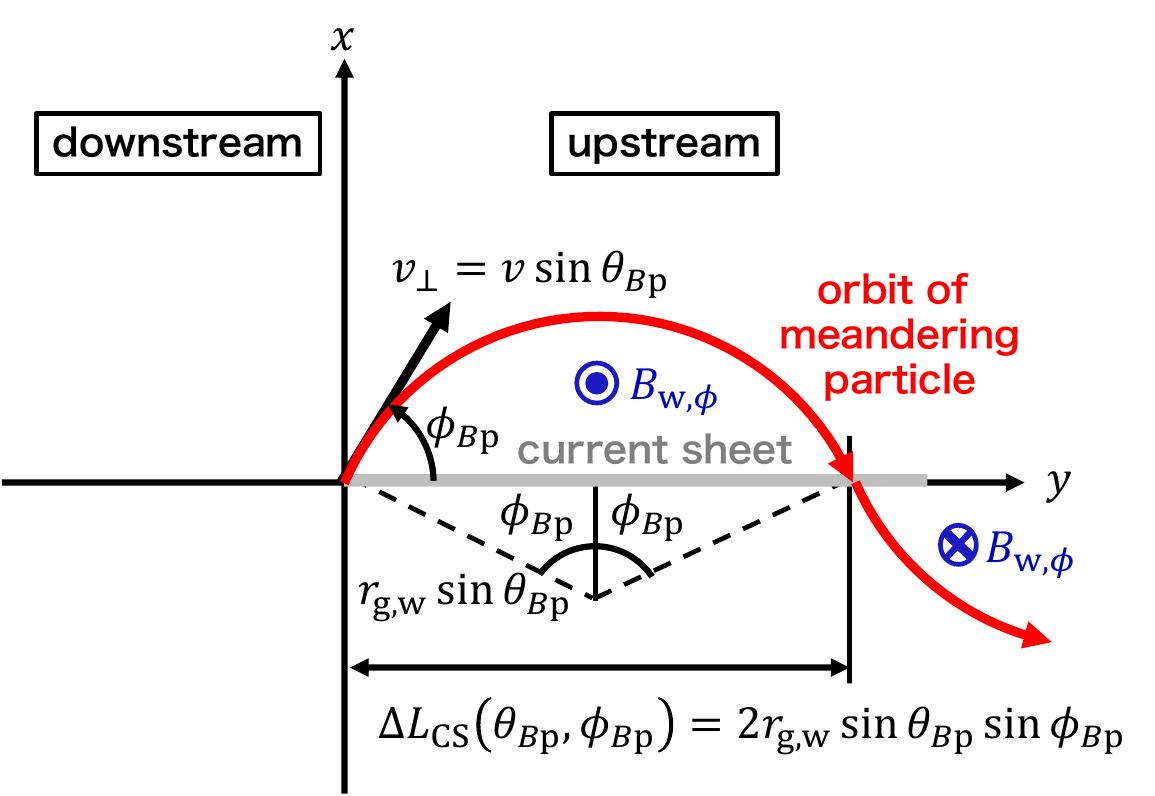}
\caption{
Particle orbit around the equator (current sheet).
The $x$ axis is the shock surface.
The positive and negative $y$ regions are the upstream and downstream regions, respectively.
The gray line is the current sheet.
The black arrow is the particle velocity perpendicular to $B_{{\rm w},\phi}$, $v_\perp$.
The red arrow is the orbit of the particle performing the meandering motion.
$\theta_{\rm Bp}$ and $\phi_{\rm Bp}$ are the pitch angle of a particle and azimuthal angle, respectively.
\label{fig:veq}}
\end{figure}
Finally, we estimate the mean particle velocity along the equator, $v_{\rm eq}$.
The particle orbit around the equator (current sheet) is shown in Fig.~\ref{fig:veq}.
The $x$ axis is the shock surface.
The positive and negative $y$ regions are the upstream and downstream regions, respectively.
The gray line is the current sheet.
The black arrow is the particle velocity perpendicular to $B_{{\rm w},\phi}$, $v_\perp$.
The red arrow is the meandering particle orbit.
$\theta_{\rm Bp}$ and $\phi_{\rm Bp}$ are the pitch angle of a particle and azimuthal angle, respectively.
Here, we consider the propagation distance, $\Delta L_{\rm CS}(\theta_{B\rm p}, \phi_{B\rm p})$, in the $y$ direction during the time when the particle is in the positive $x$ region, $\Delta t_{\rm CS}$. 
From Fig.~\ref{fig:veq}, $\Delta L_{\rm CS}(\theta_{B\rm p}, \phi_{B\rm p})$ is equal to $2r_{\rm g,w} \sin \theta_{B\rm p} \sin \phi_{B\rm p}$.
$\Delta t_{\rm CS}$ is $2\phi_{B\rm p}/\Omega_{\rm g,w}$ because $\Delta t_{\rm CS}$ is the same as the $\phi_{B\rm p}/\pi$ times the gyroperiod in the free wind region, $2\pi \Omega_{\rm g,w}^{-1}$.
Thus, the velocity of the particle moving along the equator, $v_{\rm CS}$, is given as follows:
\begin{eqnarray}
	v_{\rm CS} = \frac{\Delta L_{\rm CS}}{\Delta t_{\rm CS}} = \frac{v \sin \theta_{\rm Bp} \sin \phi_{\rm Bp}}{\phi_{\rm Bp}}. \label{eq:vcs}
\end{eqnarray}
Then, the mean particle velocity along the equator, $v_{\rm eq}$, is 
\begin{eqnarray}
	v_{\rm eq} &=& \frac{\int_{-\pi/2}^{\pi/2} d\phi_{\rm Bp} \int_0^\pi d\theta_{\rm Bp} \sin \theta_{\rm Bp} v_{\rm CS} f(\theta_{\rm Bp}, \phi_{\rm Bp})}{\int_{-\pi/2}^{\pi/2} d\phi_{\rm Bp} \int_0^\pi d\theta_{\rm Bp} \sin \theta_{\rm Bp} f(\theta_{\rm Bp}, \phi_{\rm Bp})} ~~~~~\nonumber \\
	&=& \frac{4{\rm Si}(\pi)}{3\pi} v \approx 0.79v,~~~~ \label{eq:veq}
\end{eqnarray}
where the particle distribution in the free wind region is proportional to the flux of particles that cross the shock from the downstream region to the free wind region (upstream region), $f(\theta_{\rm Bp}, \phi_{\rm Bp}) \propto v \sin \theta_{\rm Bp} \cos \phi_{\rm Bp}$ and
${\rm Si}(\pi) = \int_0^\pi dX \sin X/X \approx 1.85$.
For simplicity, we approximate all the velocities ($v_\theta, v_{\rm pl},$ and $v_{\rm eq}$) to $c/2$.

\end{document}